\documentclass[useAMS,usenatbib]{mn2e}

\usepackage{graphicx}

\title[The hierarchical growth of bulges and ellipticals]{The growth of disks and bulges during hierarchical galaxy formation. I: fast evolution \textit{vs} secular processes}

\author[C. Tonini et al.]
{C. Tonini$^{1}$
\thanks{E-mail:chiara.tonini@unimelb.edu.au},
S. J. Mutch$^{1}$,
D. J. Croton$^{2}$,
J. S. B. Wyithe$^{1}$
\\
$^{1}$School of Physics, University of Melbourne, Parkville, 3010 VIC, Australia\\
$^{2}$Centre for Astrophysics and Supercomputing, Swinburne University
of Technology, Hawthorn, VIC 3122, Australia\\
}

\begin{document}

\maketitle

\begin{abstract}

We present a theoretical model for the evolution of mass, angular momentum and size of galaxy disks and bulges, and we implement it into the semi-analytic galaxy formation code SAGE. The model follows both secular and violent evolutionary channels, including smooth accretion, disk instabilities, minor and major mergers. 
We find that the combination of our recipe with hierarchical clustering produces two distinct populations of bulges: merger-driven bulges, akin to classical bulges and ellipticals, and instability-driven bulges, akin to secular (or pseudo-)bulges. 
The model mostly reproduces the mass-size relation of gaseous and stellar disks, the evolution of the mass-size relation of ellipticals, the Faber-Jackson relation, and the magnitude-colour diagram of classical and secular bulges. 
The model predicts only a small overlap of merger-driven and instability-driven components in the same galaxy, and predicts different bulge types as a function of galaxy mass and disk fraction. Bulge type also affects the star formation rate and colour at a given luminosity. 
The model predicts a population of merger-driven red ellipticals that dominate both the low-mass and high-mass ends of the galaxy population, and span all dynamical ages; merger-driven bulges in disk galaxies are dynamically old and do not interfere with subsequent evolution of the star-forming component. Instability-driven bulges dominate the population at intermediate galaxy masses, especially thriving in massive disks. The model green valley is exclusively populated by instability-driven bulge hosts.
Through the present implementation the mass accretion history is perceivable in the galaxy structure, morphology and colours.

\end{abstract}


\section{Introduction}

The connection between dynamical processes and star formation processes in galaxies, and the origin of galaxy morphology, are fundamental pieces of the puzzle of galaxy evolution. 

In recent years, our understanding of early-type galaxies has shifted considerably. New data from surveys such as ATLAS-3D (Cappellari et al. 2011, Emsellem et al. 2011) have revealed that early-type galaxies show an unexpected complexity in dynamical features, most prominent of all their angular momentum distribution. The majority of early-type galaxies in the ATLAS-3D sample possess high rotational velocities and are indistinguishable from disks in the spin-ellipticity plane. Even the slow-rotating objects show interesting dynamical features like counter-rotating cores. The picture that emergers is one of multiple channels of formation, where secular processes and disk instability play a prominent role alongside mergers, and where gas accretion and star formation are prolonged in time. The star formation histories of massive early-type galaxies like Brightest Cluster Galaxies (BCGs) also show an increased complexity in observations, with active star formation detected down to low redshifts (see for instance Liu et al. 2012, and Oliva-Altamirano et al. 2015). 
Tonini et al. (2012) showed that hierarchical galaxy formation models can correctly predict the photometric evolution of BCGs up to the maximum redshift for which we have data for this class of objects ($z \sim 1.5$), and that the complex star formation histories produced with hierarchical clustering do not contradict the fact that these galaxies are among the oldest and reddest objects in the Universe.

In the meantime, bulges in disk galaxies have enjoyed a scrupulous investigation, and they too have emerged as a more complex and diverse class of objects than previously envisaged. Not two decades ago, the consensus was that bulges were simply smaller elliptical galaxies that were able to acquire or mantain a disk around them (Renzini, 1999). More recently it was established that bulges can be divided at least into two sub-categories. The so-called ``classical'' bulges, resembling indeed elliptical galaxies, are dynamically hot spheroids, whose structure is governed by violent relaxation during merger events, and their evolution is driven by environment (Renzini 1999). The so-called ``pseudo-bulges'' are more dynamically cold, and present intermediate features between classical bulges and disks, such as an intermediate Sersic index and velocity dispersion. Other features are definitely disky, such as flattening of the shape, a high rotation velocity and a continuity between the bulge and disk stellar populations, colours and star formation rates (Kormendy \& Kennicutt 2004, Athanassoula 2005, Kormendy \& Fisher 2008, Drory \& Fisher 2007, Peletier \& Balcells 1996, de Jong 1996, MacArthur et al. 2003). The disky features spurred Kormendy (1982) and Kormendy \& Illingworth (1982) to suggest that secular processes are responsible for the formation of pseudo-bulges, such as the funneling of gas towards the galaxy centre, while mergers cannot be responsible for their features (Fisher et al. 2009). 

The picture that is emerging is one where classifications based on morphology or global photometric properties alone cannot capture the physics of the formation of ellipticals and bulges. From the theoretical perspective, we must look at bulges and ellipticals through the mechanism of their formation, and try and predict the link between assembly history and observable properties. In particular, we need to understand the relative importance of different channels of evolution, such as mergers and secular processes (Kormendy \& Kennicutt 2004). 

If secular evolution is an important factor in shaping the bulge population, we must acquire a physical understanding of the interaction between the bulge and the disk.
Several theoretical works have investigated the formation of instabilities in disks (Krumholz \& Burkert 2010, Bournaud et al. 2011, Forbes et al. 2012, Cacciato et al. 2012) and the role of mass transfer from unstable disks in building bulges, especially at the early times of the galaxy evolution, when clumpy disks shed their mass into the galaxy centre (Noguchi 1999, Elmegreen et al. 2008, Dekel et al. 2009, Genzel et al. 2011, Forbes et al. 2014).

The next step in our theoretical understanding of disk and bulge evolution is a self-consistent picture of mass distribution and angular momentum evolution in the framework of hierarchical clustering. 
Hierarchical galaxy formation models have introduced disk instability as a source for bulge material (Croton et al. 2006, De Lucia \& Blaizot 2007, Guo et al. 2011, De Lucia et al. 2011, Fontanot et al. 2011, Henriques et al. 2015), but have never differentiated the features of the bulge based on its origin, producing instead one type of bulge, that also receives directly the material from merging satellites. A problem of theoretical models based on hierarchical clustering has been to produce galaxies with no evidence of merger-built components (Steinmetz \& Navarro 2002, Abadi et al. 2003, Kormendy \& Kennicutt 2004, Carollo et al. 2007), and to produce a realistic distribution of galaxy morphologies (Wilman et al. 2013, Fontanot et al. 2015).
In addition, models do not in general predict the size of bulges and ellipticals (with the exception of Hatton et al. 2003 and Covington et al. 2011 who use a physical recipe to calculate the size of merger remnants). Correctly producing the size of galaxies is an important test for galaxy formation models, and historically one met with scarse success. 
But galaxy size is also a fundamental physical parameter to use for the prediction of scaling relations, such as the Faber-Jackson relation, which link dynamical structure with stellar population properties. 

This paper is the first of a series in a project that aims at understanding the connection between dynamics and star formation history in hierarchical galaxy assembly, 
with the use of a semi-analytic model based on the $\rm{\Lambda}-$CDM cosmological scenario. In this work we revisit the assembly of disks and bulges, focussing on the build-up of their angular momentum and using it to characterise secular evolution $vs$ violent processes, and predict their observational signatures. We address the following questions:

$\bullet$ is hierarchical clustering able to produce the right balance between secular processes and mergers? Is this balance reflective of the properties of the merger tree, i.e. is it a direct window into the mass assembly history of the galaxy?

$\bullet$ can we produce two classes of bulges, that capture the signature observable properties of mergers and secular evolution, in particular with a more careful treatment of angular momentum evolution and mass redistribution in galaxies? 

Sections 2 presents the general features of the semi-analytic model and a summary of the new physical recipes implemented in this work for the calculation of the mass, angular momentum and size evolution of all galaxy components. Section 3 provides and overview of the different channels of galaxy evolution under study. Sections 4 and 5 present the model for the hierarchical evolution of disks and bulges respectively. Section 6 presents our results, focusing on the mass-size relations for different galaxy types, the Faber-Jackson relation, the distribution of bulge types in the galaxy population and the link with the galaxy merger history, and their predicted photometric properties. In Section 7 we discuss our findings, while in Section 8 we summarise our results. 
Throughout this paper, we adopt a value of the Hubble parameter of $h = 0.7$, and photometric magnitudes are in the AB system.

\section{Model overview}     

The skeleton of the model we produce is based on a recent incarnation of the ``Munich model'', SAGE (Semi-Analytic Galaxy Evolution, in the version presented in Lu et al. 2014; see Croton et al. 2016 for a detailed description). On this base we have built new physical recipes to follow the evolution of the galaxy star formation and dynamical structure. In particular, we have introduced the following:

$\bullet$ a prescription to follow the angular momentum evolution and mass growth of all galaxy components, including gas, stellar disk, merger-driven and instability-driven bulges;

$\bullet$ a recipe to calculate the radius of all components based on their mass and angular momentum accumulation history;

$\bullet$ a new star formation recipe, based on the evolving structure of the gaseous disk (and depending on the history of accumulation of angular momentum); 

$\bullet$ a spectrophotometric model to calculate galaxy luminosities and build mock galaxy catalogues.

The approach of this method is to evolve galaxy properties following each hierarchical clustering event (gas cooling and mass accretion in the form of gas and stars), based on the existing galaxy history. New gas accreted by the galaxy will settle onto the existing disk, incrementally changing its angular momentum; this in turn will have an effect on the density profile, which will determine how much gas can be turned into stars. An incoming satellite will be absorbed into the galaxy and modify its structural parameters depending not only on the satellite's properties, but also on the galaxy's current mass and angular momentum distribution (i.e. its morphology).

\subsection{The basic semi-analytic model}

We use numerical N-body simulations to construct the dark matter structure where galaxy formation is nested; for this work we choose the combination of box size and resolution of the Millennium simulation (Springel et al. 2005). We use the merger trees, that record the assembly history of every object in the box, obtained by Springel et al. (2005) with the L-HALOTREE algorithm, after structures were identified with SUBFIND (Springel et al. 2001)

The SAGE model populates each dark matter halo at the beginning of a merger tree with the cosmic fractional amount of baryons in the form of hydrogen, and then follows the cooling of the gas, the star formation rate, the feedback from supernovae and active galactic nuclei, so that at each timestep from the first redshift down to $z=0$ the evolution of the mass and metallicity of stars, hot and cold gas is accounted for. At the same time, the merger tree provides the hierarchical assembly, i.e. the accretion of substructures. The model follows each satellite as it enters the central halo, (where it is subjected to dynamical friction, ram pressure stripping, tidal torques) and eventually merges the satellite with the central galaxy or distrupts it into the central halo, forming an intra-cluster component. Each of these processes is described in Croton et al. (2016; see also Lu et al. 2014, Croton et al. 2006).

The model outputs the physical properties of the galaxy and its star formation history (i.e. a record of all the stellar populations that the galaxy contains, whether they were formed in-situ or accreted from satellites). The star formation history is then used in post-processing by a spectro-photometric model, which produces galaxy spectra and luminosities in any desired photometric band (Tonini et al. 2009, 2010, 2012). 

In what follows we present our new physical recipes and implementations, which have been applied to the basic version of SAGE (Lu et al. 2014; Croton et al. 2016).   

\section{The channels of galaxy evolution}  
\label{sec:channels}

In the model all galaxies start out as disks. The initial condensation of baryons in the centre of the dark matter halo forms a disk of cold gas, where stars form in dynamical equilibrium with it. 
Depending on the environment of this galaxy, or in other words depending on this galaxy's merger tree, the galaxy evolution and mass growth can proceeed through a combination of four main channels. The relative importance of these channels is going to determine the galaxy dynamical structure, star formation history and morphology. 

\smallskip

$\textbf{1.}$ Smooth accretion of gas in an undisturbed environment (Section 4): 

\noindent This is the steady cooling of gas from the hot component gravitationally trapped in the dark matter halo or cooling flows of moderate intensity. The hot gas reservoir is composed of 
gas that is accreted by the halo and shock-heated at the halo virial temperature. In addition it contains gas reheated from the galaxy's feedback mechanisms. Quiescent accretion leads to the formation of a disk, and a steady inside-out stellar disk growth regulated by local star formation and stellar feedback (see for instance Guo et al. 2011, Croton et al. 2016, Lu et al. 2014, Hatton et al. 2003). 

\smallskip

$\textbf{2.}$  Disk perturbations from chaotic gas accretion (Sections 4.2, 5.1): 

\noindent When the gas cooling times are short compared to the halo dynamical time, like in the case of cooling flows with high infall rates or gas-rich minor mergers, the central galaxy disk receives a large surplus of gas in a short time, and thus becomes gravitationally unstable.
To regain dynamical equilibrium, the disk must rearrange its mass distribution, through angular momentum dissipation (in the gas) and angular momentum transfer (in the stars). We assume that the excess gas sinks to the galaxy centre and condenses rapidly in a violent burst of star formation, and depending on the severity of the upset, a fraction of the disk stellar mass loses angular momentum and migrates to the galaxy centre (see Dutton et al. 2007, Croton et al. 2016, 2006). 

In this model the accumulation of stars in the galaxy centre leads to the growth of an \textbf{instability-driven bulge} (column 1 of Fig. 1). This bulge is composed of disk material and newly-formed stars, and retains a memory of the dynamical state of the disk. Its angular momentum is aligned with that of the disk, its mass distribution is flattened in the disk direction, it rotates with the same velocity, and it exhibits similar stellar populations. 

\smallskip

$\textbf{3.}$ Minor mergers (Sections 5.1, 5.2):

\noindent Satellites that survive the disruptive forces of the dark matter halo impact the central galaxy and trigger a structural evolution that depends on the galaxy mass distribution (columns 2, 3, 4 of Fig. 1). 

We assume that the dominant dynamical component of the central galaxy regulates the mass deposition. In the case of an elliptical galaxy, the dominant mass component is spheroidal and the newly accreted stellar mass accumulates in shells around it, thus growing the \textbf{merger-driven bulge}. The growth of its radius depends on the incoming satellite mass and orbital parameters. 

In the case of galaxies dynamically dominated by a disky mass component (stellar disk or instability-driven bulge), the mass deposition likely happens on the plane of the disk, which offers a much wider cross-section for impact than the bulge (as is also seen in hydro-dynamical simulations, for instance in Abadi et al. 2003b). 
The gravitational imbalance triggers disk instabilities, which again cause angular momentum transfer and the radial migration of stars to the galaxy centre, where they grow the \textbf{instability-driven bulge}. Regardless of the central galaxy morphology, if the satellites contains gas, it dissipates its angular momentum and it sinks to the galaxy centre, where it is bursted into stars that add to the instability-driven bulge.

\smallskip

$\textbf{4.}$ Major mergers (Section 5.2.3):

\noindent When two galaxies of similar mass (with a merger ratio around $1:3$ or larger) collide, a new \textbf{merger-driven bulge} is formed (column 5 of Fig. 1). 
The interaction is governed by violent relaxation, so that the final equilibrium configuration of the galaxy depends only on the overall gravitational potential, and all memory of the initial configuration of the two progenitors (incoming galaxies) is lost (see Binney \& Tremaine 2008). We assume this object is completely pressure-supported. 
Major mergers typically compress the gas in the two incoming progenitors, triggering a violent burst of star formation. At the same time, gas can be funnelled to the centre of each of the two systems just before merging, triggering AGN activity in one or both.

\medskip

We point out that, for each event in the galaxy's history, the model treats the mass accretion depending on the \textit{instantaneous} galaxy structure and the angular momentum of the encounter. However it is hierarchical clustering that determines the galaxy properties; the merger tree is entirely responsible for the frequency and magnitude of each type of event. By keeping track of both mass and angular momentum, the model allows the assembly history to leave its imprint on the galaxy observables.

In the model a galaxy can be composed of one or more components: a disk, an instability-driven bulge, a merger-driven bulge. Any long-lived cold gas (i.e. gas that has time to settle into the galaxy without being immediately turned into stars) will be found in the gaseous disk. Bulges only contain gas temporarily, as a result of instabilities, gas-rich minor mergers or when ejected from their own supernovae. 
When the stellar mass is mostly in the disk component, we will refer to the object as a disk galaxy. When instead the bulge dominates the stellar mass budget, if the instability-driven bulge dominates, the object is akin to a lenticular (S0) galaxy, while if the merger-driven bulge dominates, the object is an elliptical galaxy.

\subsection{A word about nomenclature}

In the literature, ``spheroids'' (both bulges and elliptical galaxies) in general are often characterised by their surface brightness density profile, or Sersic index. Typically ``classical bulges'' are defined as objects with a Sersic index of 4 (or in other words, a deVaucouleurs surface brightness density profile), and ``secular bulges'' are defined as objects with a lower Sersic index, peaking around 2 (Fisher \& Drory 2008, Kormendy \& Kennicutt 2004). 

Note that another definition is that of ``early type'' object, which is even broader. This definition can be based for instance on visual inspection, photometric properties or surface brightness density fits, and in all cases it can include both ``classical'' and ``secular'' bulges, or in some cases even non star-forming disks.

Although the surface brightness density profile is linked to the mechanism of formation, there is a certain degree of ambiguity in these definitions, mainly because they imply a modeling of light profiles which may or may not mirror the galaxy structure in an obvious way. 

In this work, we distinguish galaxy components through their mechanism of formation, and \textit{then} predict their observational properties. For this reason, for bulges in particular we avoid the nomenclature of classical and secular, as well as that of ``pseudo-bulge''. The semi-analytic model does not calculate full mass profiles for all galaxies (and therefore, no surface brightness density profiles either), but only the fundamental parameters of the galaxy components (in this case, masses and half-mass radii). In other words, the Sersic index of the galaxy mass components is left as an unknown. To predict \textit{some} observational properties, such as the galaxy velocity dispersion used to build the Faber-Jackson relation, we need to assume a Sersic index and calculate the full \textit{mass} density profile in post-processing. When this is the case, it will be clearly indicated.

\section{The formation and evolution of disks}

We cannot correctly model bulges if we do not correctly model disks first. Bulges either form from disk material, or from the merger of galaxies that in turn contain, or were shaped from, disks. The disk structure affects the bulge structure in subtle and important ways, such as the mass-size relation, the star formation history (and therefore the composition of the stellar populations) and the metallicity. These connections are explored in a future paper (Tonini et al., in prep). In this Section we present our new physical recipe for the evolution of the disk angular momentum and size, based on the disk mass accretion history. The resulting structural properties of disks serve as a base for the build-up of bulges. 

When a halo is identified by the halo finder for the first time, its mass, spin and virial radius are calculated. The halo attracts hydrogen (by a fractional amount corresponding to the cosmic average, $~17 \%$ of the halo mass; see Lu et al. 2014), and this gas is initially assumed to be dynamically coupled with the halo. The gas cools radiatively and sinks to the centre of the halo, conserving its specific angular momentum, and settling into a disk. The initial disk mass is determined by the cooling rate, its initial spin matches that of the halo, and the disk scale radius is directly proportional to the halo spin parameter and virial radius through the Mo et al. (1998) relation.

However we note that the Mo et al. (1998) prescription for the disk radius may not be applicable after a baryonic component is established in the centre of the halo. The halo itself is subject to tidal interactions with its neighbours, and its spin can change quite dramatically and quite fast. These changes in angular momentum are driven by the torques in the outer regions of the halo. The baryonic disk however
is much denser and more tightly bound than the dark matter, and should remain stable against these sudden perturbations, safely nestled in the innermost deepest part of the gravitational potential well.

Crucially, the dynamical state of the disk and the star formation rate are closely coupled. 
Variations in disk size are necessarily accompanied by variations in gas density, which determines the star formation rate directly (Croton et al. 2016, 2006). If the disk size is directly governed by the halo spin, this results in wildly varying disk densities, and an artificially bursty star formation history, with consequences over the galaxy metallicity and photometric properties as well as the mass and size (Tonini et al. in prep). 

In this work, beyond the initial collapse of the gas to form the proto-galaxy, during which the dark matter halo properties are imprinted on the disk, we assume that the dynamical coupling of the galaxy with the halo becomes marginal. The disk (and galaxy) dynamical structure is instead determined by
the string of processes that it undergoes during its lifetime. These include cooling of gas from the hot gas component trapped the dark matter halo, accretion of gas from infalling satellites, transformation of gas into stars and removal of gas due to feedback, and gravitational instabilities. At each of these events, the disk adjusts to a new equilibrium configuration, and the mass, angular momentum and the radial extension of both the gaseous and stellar components evolve.

\subsection{The gaseous disk}

The mass and angular momentum of the cold gas disk are the result of the accumulation or loss of material over time, caused by cooling of the hot gas component trapped in the dark matter halo, star formation, and incoming satellites. Following Guo et al. (2011), the variation in mass and angular momentum of the gaseous disk can be calculated as: 
\begin{eqnarray}
\delta M_{\rm{gas}} & = & \dot{M}_{\rm{cool}} \delta t - \dot{M}_{*} \delta t + M_{\rm{sat, gas}}~,  \\
\delta \vec{J}_{\rm{gas}} & = & \delta \vec{J}_{\rm{gas, cooling}} + \delta \vec{J}_{\rm{gas,sat}} + \delta \vec{J}_{\rm{gas,SF}} ~.
\label{mass_and_j}
\end{eqnarray}

The gas which is cooling from the hot component is assumed to be in dynamical equilibrium with the dark matter halo, so that at the moment of accretion it shares the halo specific angular momentum. The total incremental variation of the cold gaseous disk angular momentum is therefore:
\begin{equation}
\delta \vec{J}_{\rm{gas, cooling}} = \dot{M}_{\rm{cool}} \frac{\vec{J_{\rm{DM}}}}{M_{\rm{DM}}} \delta t ~,
\label{jgas_cooling}
\end{equation} 
where $\dot{M}_{\rm{cool}}$ is the cooling rate (for details see Lu et al. 2014). 
$\delta \vec{J}_{\rm{gas, cooling}}$ is aligned with the instantaneous halo spin.

The gas that is added to the disk as a result of an accretion of a satellite is also assumed to be in dynamical equilibrium with the dark matter halo, so that its contribution to the total angular momentum of the cold gaseous disk is: 
\begin{equation}
\delta \vec{J}_{\rm{gas,sat}} = M_{\rm{sat, gas}} \frac{\vec{J_{\rm{DM}}}}{M_{\rm{DM}}}~.
\label{jgas_sat}
\end{equation} 
Notice that this is strictly correct in the case where the gas from the satellite is stripped by a process like ram-pressure stripping, thus decoupling from the satellite's orbit and settling into the surrounding dark matter halo before being accreted by the disk. This is not always the case, and to various degrees the gas could retain a ``memory'' of the satellite orbit, which would alter its angular momentum contribution. However, this would only be significant for relatively massive satellites in isolated events. We assume that, over the lifetime of a galaxy, multiple accretion events due to minor mergers are likely to average out the effects of single orbits, and the overall effect is dominated by the spin of the dark matter halo. Therefore $\delta \vec{J}_{\rm{gas,sat}}$ is aligned with the instantaneous halo spin. 

Gas is removed from the gaseous component when star formation occurs. Stars are assumed to form in dynamical equilibrium with the gas, and therefore they carry the gas specific angular momentum. The incremental variation of the angular momentum in the gas is equal and opposite to the one in the stellar disk, and is aligned with the instantaneous gas disk angular momentum.   
\begin{equation}
\delta \vec{J}_{\rm{gas,SF}} = - \delta \vec{J}_{*}~.
\label{delta_jstars}
\end{equation} 
This quantity will be described in the next Subsection. 

Note that variations in the dark matter halo angular momentum will propagate down into the disk angular momentum, but only when the gas disk acquires new mass. Every variation in $\vec{J_{\rm{DM}}}$ will be weighted by the ratio $\delta M_{\rm{gas}}/M_{\rm{DM}}$, thus producing a much more stable evolution of the gas angular momentum compared to a Mo et al. (1998) implementation. 
Note also, that while the increment $\delta \vec{J}_{\rm{gas}}$ is temporarily aligned with the instantaneous halo spin and the increment $\delta \vec{J}_{*}$ is temporarily aligned with the instantaneous gas angular momentum, the total $\vec{J}_{\rm{DM}}, \vec{J}_{*}, \vec{J}_{\rm{gas}}$ have been built over time and can be misaligned. 

To calculate the gas disk radius, we adopt the standard assumption (see Guo et al. 2011) that the galaxy disk components (cold gas and stellar disks) are thin, centrifugally supported, and characterised by exponential surface density profiles of the type 
\begin{equation}
\Sigma(r) = \Sigma_{\rm{0}} \ \rm{exp}(-r/R_{\rm{D,gas}}) ~, 
\label{gas-density}
\end{equation}
where $R_{\rm{D,gas}}$ is the characteristic scale-length, and the central density is $\Sigma_{\rm{0}} = M_{\rm{gas}}/(2 \pi R_{\rm{D,gas}}^2 )$. In the general scenario where the dark matter halo dominates the galaxy rotation curve everywhere but at the very centre of the galaxy, the scale length of the gaseous disk is given at any time by: 

\begin{equation}
R_{\rm{D,gas}} = \frac{J_{\rm{gas}} / M_{\rm{gas}}}{2 V_{\rm{max}}}~,
\label{Rgas}
\end{equation} 
where $V_{\rm{max}}$ is the peak rotation velocity of the dark matter halo (Guo et al. 2011). Notice that an exact solution would require the use of the full rotation curve of the galaxy, calculated from the mass density profiles of all galaxy components.
However in observed disk galaxies baryons and dark matter conspire to produce flat rotation curves at a radius equal to twice the disk scale length, with a velocity that is well approximated by the halo peak velocity. 

The gas disk radius changes whenever the angular momentum $J_{\rm{gas}} $ evolves, but also every time it acquires or loses mass at fixed angular momentum. For example if stellar feedback expels a fraction of the gas, the remaining material needs to redistribute its angular momentum, and $\delta R_{\rm{D,gas}} \propto 1/ \delta M_{\rm{gas}}$.

\begin{figure*}
\includegraphics[scale=0.7]{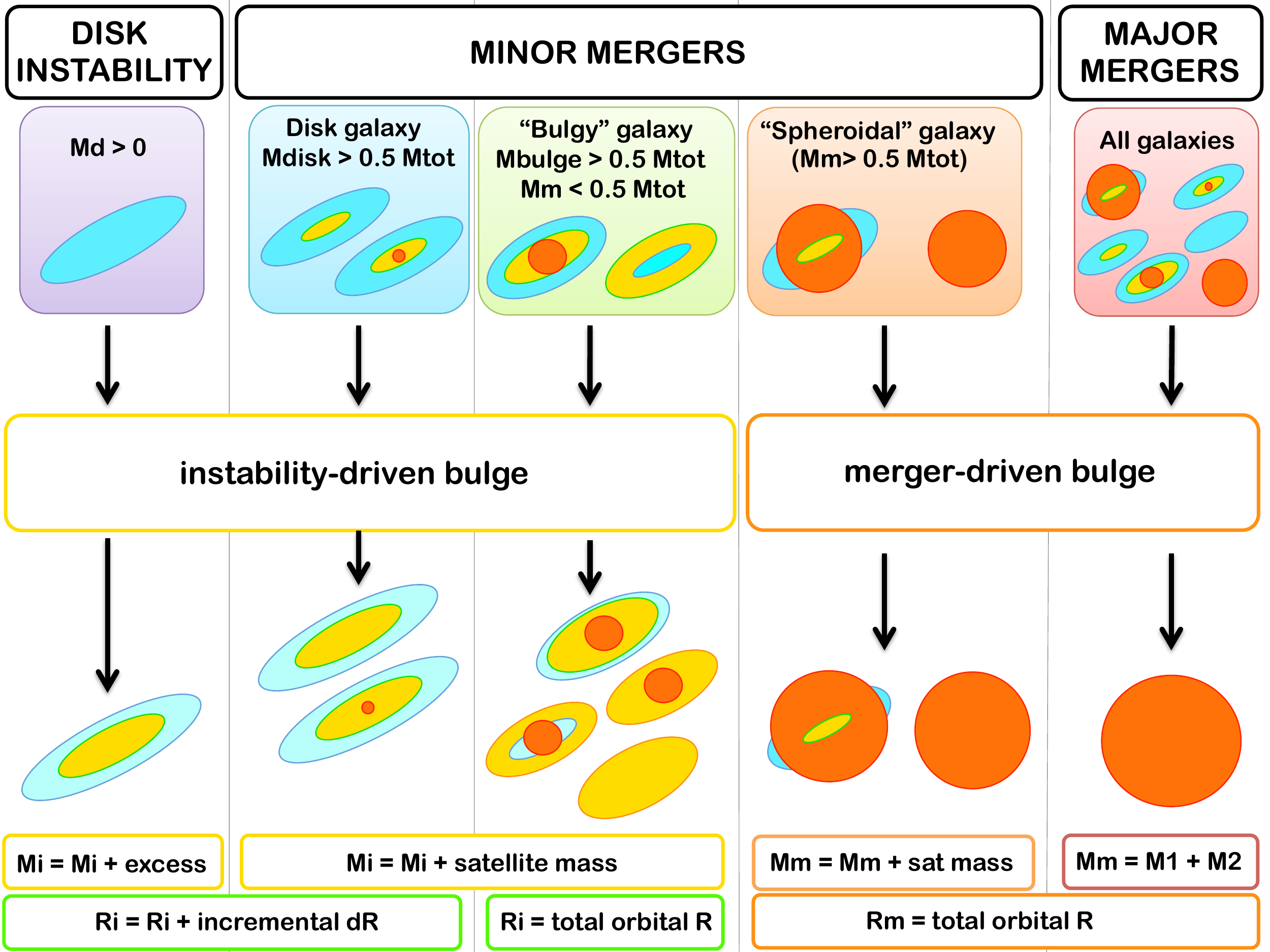}
\caption{The formation and evolution of bulges. \textbf{Md}: total disk mass. \textbf{Mtot}: total stellar mass. \textbf{Mdisk}: stellar disk mass. \textbf{Mbulge}: total bulge mass. \textbf{Mi}: stellar mass of the instability-driven bulge. \textbf{Mm}: stellar mass of the merger-driven bulge. \textbf{Ri}: half-mass radius of the instability-driven bulge. \textbf{Rm}: half-mass radius of the merger-driven bulge. \textbf{M1} and \textbf{M2}: mass of the major merger progenitors. The \textbf{incremental dR} refers to the radius defined in Eq.(15, 16), while \textbf{total orbital R} refers to the radius defined in Eq.(17).}
\label{cartoon}
\end{figure*}

\subsection{Star formation and the stellar disk}

Since we follow the evolution of the disk gas angular momentum, we know at any time the density of the gas disk. 
We can take advantage of this knowledge to produce a star formation law that depends on the disk density profile, and therefore on the angular momentum itself. This star formation law is different from previous prescriptions in the literature (such as Lu et al. 2014, Croton et al. 2016, 2006, Guo et al. 2011, Henriques et al. 2015, Hatton et al. 2003, Baugh et al. 2005, Bower et al. 2006, Monaco et al. 2007). 

The gas density threshold for star formation, as constrained by observations (see Kormendy \& Kennicutt 2004 and references therein), is $\Sigma_{\rm{crit}} = 10 M_{\odot}/pc^2$. Given Eq.(\ref{gas-density}), we can calculate the radius $R_{\rm{crit}}$ at which the gas density profile drops below $\Sigma_{\rm{crit}}$ as: 
\begin{equation}
R_{\rm{crit}} = R_{\rm{D,gas}} \cdot \rm{Log} \left( \frac{\Sigma_{\rm{0}}}{\Sigma_{\rm{crit}}} \right)~.
\label{Rcrit}
\end{equation}
The mass inside this radius is
\begin{equation}
M_{\rm{crit}} = M_{\rm{gas}}  \left[ 1 - \exp{ \frac{R_{\rm{crit}}}{R_{\rm{D,gas}}}} \cdot \left( 1 + \frac{R_{\rm{crit}}}{R_{\rm{D,gas}}}  \right)  \right]~,
\label{Mcrit}
\end{equation} 
and is converted into stars over a dynamical time $t_{\rm{dyn}} = R_{\rm{crit}} / V_{\rm{vir}}$ (where $V_{\rm{vir}}$ is the halo virial velocity), with an free efficiency parameter $\epsilon = 0.25$.
The star formation rate is therefore defined as: 
\begin{equation}
SFR = \epsilon \cdot M_{\rm{crit}} / t_{\rm{dyn}}~.
\label{SFR}
\end{equation} 
where $R$ is the fractional amount of the new stellar mass that is instantaneously recycled back to the gas component (we set $R=0.43$ as in Lu et al. 2014, Croton et al. 2016). 

Note that $M_{\rm{crit}}$ and $R_{\rm{crit}}$ are functions of the disk scale-length and therefore the star formation rate is a function of $\vec{J}_{\rm{gas}}$; given the same disk mass, to a higher angular momentum corresponds a lower star formation rate. 

With the increment in stellar mass $\delta M_{\rm{stars}} = \epsilon \cdot M_{\rm{crit}}$, the stellar disk acquires  angular momentum $\delta \vec{J}_{*}$ (see Eq.~\ref{delta_jstars}). The stars are formed in dynamical equilibrium with the gas, but only the innermost part of the gas disk is turned into stars, i.e. inside the critical radius. Therefore the stars will acquire an angular momentum at most equal to that of the gas at the critical radius. We calculate it as:
\begin{equation}
\delta \vec{J}_{*} = \delta M_{\rm{stars}} \cdot 2 V_{\rm{max}} R_{\rm{crit}} \frac{ \vec{J}_{\rm{gas}} }{ |\vec{J}_{\rm{gas}} |}~.
\label{jstars}
\end{equation} 

The stellar disk immediately loses a fraction $R$ of its newly formed stars $\delta M_{\rm{stars}}$, exploding as supernovae on timescales comparable to the simulation's timestep, and is recycled back into the gas component (we set $R=0.43$ as in Lu et al. 2014, Croton et al. 2016). 
The stellar disk characteristic scale-length, assuming an exponential density profile, can be defined as: 
\begin{equation}
R_{\rm{D}} = \frac{J_{*} / M_{\rm{D}}}{2 V_{\rm{max}}}~,
\label{Rdisk-temp}
\end{equation} 
where $M_{\rm{D}}$ is the mass of the stellar disk. 

Note that $\delta \vec{J}$ is a vector, which implies that both the gaseous and the stellar disk are allowed to both expand or shrink in radius at a fixed or growing mass, thus decreasing or increasing their density. 
In addition, notice again that the stellar disk radius varies not only when the disk angular momentum varies, but also every time the disk acquires or loses stellar mass at fixed angulat momentum. The most significant example of this process is represented by gravitational instabilities.

When accreting new material (whether in the form of gas or satellites), the stellar $+$ gaseous disk can climb above a critical mass that pushes it out of dynamical equilibrium; this condition is described in Croton et al. (2006; see also Kormendy \& Kennicutt 2004) as:  
\begin{equation}
M_{\rm{disk}} \geq \frac{V_{\rm{C}}^2 R_{\rm{disk}}}{G }~, 
\label{instability}
\end{equation}
where $V_{\rm{C}}$ is the disk circular velocity. Here we take $ R_{\rm{disk}}$ as the mass-averaged scale length of the stellar $+$ gaseous disk, and $M_{\rm{disk}}$ as the total disk mass.
At this point the disk, too heavy for its own rotation velocity, must find a new dynamical equilibrium state. We assume that the excess mass is transported inwards, along the disk towards the galaxy centre, while angular momentum flows outwards.  
The excess gas sinks towards the denser galaxy centre and is consumed in a burst of star formation.
The excess stellar material is shedded from the disk and sinks to the galaxy centre. Both the stars created in the bursts and the ones transferred from the disk end up in the instability-driven bulge (see next Section).

If we assume angular momentum conservation in both the gaseous and the stellar disk while $M_{\rm{excess, i}}$ sinks into the galaxy centre, then the angular momenta of each component $i$ must be redistributed in what remains of each disk, which now have a mass $M_{\rm{i}} - M_{\rm{excess, i}}$. 
For a self-gravitating system such as a disk, which has negative specific heat, it is energetically favourable to expand the outer parts in response to the increase in central density (Lynden-Bell \& Wood 1968, Lynden-Bell \& Kalnajs 1972, Tremaine 1989, Binney \& Tremaine 2008).
While the total angular momentum is conserved, a decrease in mass causes the specific angular momentum of the disk to increase, with the consequence that its characteristic radius also increases, by an amount given by: 
\begin{equation}
\delta R_{\rm{D,i}} = \frac{J_{\rm{i}} / (M_{\rm{i}}-M_{\rm{excess,i}})}{2 V_{\rm{max}}}~.
\label{Rdisk-final}
\end{equation}
This prescription is similar in spirit to the post-processing recipe to calculate the expansion of the disk radius in response to bulge formation described in Dutton et al. (2007), with the difference that here we calculate the angular momentum variation of the disk self-consistently (rather than in post-processing), step by step at every episode of the galaxy history. 

In the model, disk instabilities that cause the internal evolution of the disk are caused by external perturbations, such as mass accretion and mergers with satellites. These depend on the richness of the merger tree, or in other words, the density of the galaxy's environment.

\section{The formation and evolution of bulges}

Three of the four channels of galaxy evolution described in Section~\ref{sec:channels} lead to the growth of bulges, as illustrated in Fig.(1). 
Our implementation of bulge formation and evolution differs from that of other models in the literature (such as Lu et al. 2014, Croton et al. 2016, 2006, Guo et al. 2011, Henriques et al. 2015, Hatton et al. 2003, Baugh et al. 2005, Bower et al. 2006, Monaco et al. 2007, De Lucia \& Blaizot 2007).
The bulge physical properties at any time depend on 1) the current channel of mass accretion, and 2) the current galaxy properties, which are determined by its mass accretion history and angular momentum evolution. 

\subsection{Disk instabilities and the bulge}

When the disk becomes gravitationally unstable, we assume that the excess mass $M_{\rm{unstable}}$ is transferred to the galaxy centre, and accumulates into an \textbf{instability-driven bulge}. If the excess mass includes gas, this is immediately turned into stars, with the current metallicity of the gas itself. The transferred stellar component on the other hand retains the metallicity of the disk from where it originated. 

The instability-driven bulge not only inherits some of the stellar populations of the disk, but we assume it retains some memory of the instantaneous disk dynamical structure. The material that composes this bulge used to be rotationally supported, and by loss of angular momentum it has sunk to the galaxy centre, along the plane of the disk. The result is an object with 
an angular momentum vector aligned with that of the stellar disk and a fraction of the disk rotation. The amount of angular momentum dissipated in this process is undetermined, but observations can constrain it. Fisher \& Drory (2008, 2010) show that bulges grown from instabilities present a median half-mass radius that correlates with the disk scale-length as $R_{\rm{bulge}} = 0.2 R_{\rm{D}}$, with a significant scatter around this relation. In other words, at fixed rotation velocity stars in the instability-driven bulge have retained $\sim 20\%$  of their initial specific angular momentum. Their findings also suggest that the size of bulges formed through instability processes evolve their size slowly as they grow in mass, and increase their density. 

We build a recipe for the evolution of the radius of the instability-driven bulge that depends on the relative magnitude of the perturbation to the bulge structure, and on the current structure of the disk. If $\delta M$ is the mass that is being transferred to the bulge, and $R_{\rm{D}}$ is the scale-length of the stellar disk, the bulge radius after the instability event is:
\begin{equation}
R_{\rm{i}} = R_{\rm{i, OLD}} + \delta R = \left ( \frac{R_{\rm{i, OLD}}  M_{\rm{i, OLD}}  + \delta M \cdot 0.2 \cdot R_{\rm{D}} }{M_{\rm{i, OLD}} + \delta M} \right) 
\label{r_instability}
\end{equation}
where $R_{\rm{i, OLD}} $ and $M_{\rm{i, OLD}}$ are the radius and mass of the instability-driven bulge before the instability event. This channel of evolution is depicted in Fig.(\ref{cartoon}) on the far left, where Eq.(\ref{r_instability}) is referred to as \textit{incremental dR}. 

Since the radius of the disk is allowed to shrink during its evolution, this implies that the radius of the instability-driven bulge can also shrink, while its mass grows. 
In fact this becomes more and more common as the bulge becomes more massive compared to the disk. It is also possible that a disk can entirely disappear into the instability-driven bulge through this mechanism. 
This recipe implies that, the more mature an instability-driven bulge is, the denser it becomes. 
 
\subsection{Mergers and the two bulges}

There is a long-standing debate about the origin of bulges and the evolution of elliptical galaxies. From the theoretical point of view, mergers are the natural culprits for the formation of non-disky components in galaxies, but some of the observed bulge properties, such as the high metallicity and old stellar populations, seem hard to reconcile with the merger scenario. In addition, recently bulges have emerged as a complex and diverse population of objects (see Kormendy \& Kennicutt 2004, Fisher \& Drory 2008, 2010). 

In this work, we move away from the classical implementation of mergers found in SAGE and semi-analytic models in general, and focus on the way \textbf{baryons} interact during a merger. Traditionally, it is enough for a satellite to reach the centre of the central halo to be absorbed and its material be deposited in the bulge. However here we take into account the fact that, after sinking deep into the relatively low-density dissipationless halo, reaching the centre of the \textbf{galaxy} is a very different matter. In fact, observations of galaxies across different morphological types from spirals to giant ellipticals show that the disruption of satellites produce arcs and shells of debris in the outskirts of the central galaxy. The satellite responds to the shape of the gravitational potential well of the central galaxy long before it can reach the centre (Abadi et al. 2003). 

The operating scenario for our model is that the final trajectory and disintegration of the satellite depends on the dominant mass component and structure of the central galaxy. This implies that the fate of the satellite and the evolution of the galaxy following its absorption will depend on the current galaxy structure itself (i.e. its morphology).
By following the evolution of the galaxy angular momentum and mass, we can treat a merger as any other perturbation, and we have a recipe to calculate the galaxy response to the event. The larger the perturbation, the more dramatic the galaxy evolution will be. At the far end of the spectrum, major mergers completely destroy the structure of the two progenitors. We adopt the threshold $M_{\rm{satellite}}/M_{\rm{central}} = 0.3$ as the divide between major and minor mergers. 

We first need to define which galaxy is the central and which one is the satellite. This is usually done by comparing the mass of the two dark matter halos participating in the collision, as is done in SAGE for instance. This becomes problematic when the baryon-to-dark matter mass ratio deviates from the average. Notice that in SAGE, and any other model where  no attempt is made to calculate the angular momentum or radius of galaxies, every collision is treated as a symmetrical event. Two masses go in, one mass comes out. But if any property of the galaxy depends on the galaxy history and is not additive (like when the angular momentum or the radius are results of incremental variations), this approach can no longer be adopted. Moreover if the dynamics of the collision is dominated by the baryons, we need to propagate down the merger tree the properties of the galaxy that plays the role of central in the collision. So we define the central galaxy as the one with the largest stellar $+$ cold gas mass. In some cases (less than $10 \%$), this is the one with the smaller halo. This galaxy will then ``occupy'' the other's larger halo, and its own properties will be propagated down the merger tree.

\subsubsection{Minor mergers on disks}

In a spiral galaxy the mass component that dominates the dynamics of the encounter is the  disk. We select such objects to have $M_{\rm{disk}} \geq 0.5 M_{\rm{star}}$ (second column of Fig.~\ref{cartoon}). If the impact is face-on, the satellite has a higher probability of impacting the disk at a radius $r \gg 0$, rather than the bulge. If the impact is not face on, this probability increases up to 1 for an edge-on collision. 

The satellite mass (gas and stars) is thus added to the galaxy disk component, and this triggers a disk instability. The satellite gas sinks to the centre and is consumed in a burst of star formation (as in Lu et al. 2014, Croton et al. 2016), that adds stellar material to the \textbf{instability-driven bulge}. We assume that the stellar disk acquires the satellite stars $M_{\rm{sat,*}}$, and the gravitational instability ripples across the disk in the radial direction, causing stars from the inner disk to sink into the bulge, with the shedded mass being $M_{\rm{excess}} = \gamma M_{\rm{sat,*}}$. 

The collision is filtered by the disk, and instability-driven bulge grows in radius incrementally depending on the disk instantaneous structure, in analogy with Eq.(\ref{r_instability}):

\begin{equation}
R_{\rm{i}} = \left ( \frac{R_{\rm{i, OLD}}  M_{\rm{i, OLD}}  + \gamma M_{\rm{sat,*}} \cdot 0.2 \cdot R_{\rm{D}} }{M_{\rm{i, OLD}} + \gamma M_{\rm{sat,*}}} \right)~, 
\label{r_instability_merger}
\end{equation}
where $R_{\rm{i, OLD}} $ and $M_{\rm{i, OLD}}$ are the radius and mass of the instability-driven bulge before the collision. The quantity $\gamma$ is treated as a free parameter. In effect, it depends on the details of the encounter and the structure of the central disk. At first order, $\gamma \sim 1$, which is the value we adopt here. In future work we will explore the effects of different choices for this parameter. 

Notice that, even if the end result is adding a mass equal to $\gamma M_{\rm{sat,*}}$ to the bulge, the stellar populations that are locked in the instability-driven bulge come from the inner disk. This is consistent with observations that show that these kind of bulges present a continuity of stellar ages and metallicities with the disk, and a similar rotation. 

\subsubsection{Minor mergers on bulgy galaxies}

This case represents galaxies defined as: $M_{\rm{bulge}} > 0.5 M_{\rm{star}}$ \textbf{and} $M_{\rm{m}} < 0.5 M_{\rm{star}}$, i.e. the bulge dominates the mass, but the merger-driven bulge $M_{\rm{m}} $ accounts for less than half the stellar mass (third column of Fig.~1). These galaxies are rare, accounting for a few percent of the total population. They represent two categories of galaxies: 1) a few true hybrids, with a disk, an instability-driven bulge and a merger-driven bulge of comparable masses, but mostly 2) 
a dominating instability-driven bulge plus a small disk and a small old merger-driven bulge. These galaxies are neither purely disky nor spheroidal, however they are flat and have a well defined spin and rotation. 

In this case the disk is not the dynamically dominant mass component (in general $M_{\rm{disk}} << 0.5 M_{\rm{star}}$), thus we assume it is no longer able to regulate the encounter and absorb the satellite. The simulation does not record the trajectory of the satellite after it enters the virial radius of the central halo, so the semi-analytic model does not have information on the location of the impact. But we know the total bulge is dominating the mass,
and the instability-driven bulge has in general a larger surface than the merger-driven one, which in fact would be nested in its centre. Therefore we generalise this situation by assuming that the satellite is absorbed by the \textbf{instability-driven bulge} which grows its mass by $M_{\rm{sat,*}}$, plus the stellar material originating from a star formation burst of the gas contained in the satellite. 
If present, we assume that the merger-driven bulge remains at the centre unperturbed, as we do not attempt to model mass transfer between the bulges. 

The radius of the instability-driven bulge can no longer be regulated by the disk, which in most cases is very small. Rather, the material will deposit itself in shells (or in this case, rings) at the periphery of the bulge, depending on the satellite orbital parameters. We calculate the radius with the \textit{orbital R} recipe, described in the next Subsection.

\subsubsection{Mergers on spheroids and major mergers} 

The last two columns of Fig.(\ref{cartoon}) show the formation and growth of the \textbf{merger-driven bulge.} This scenario applies to all major mergers regardless of the properties of the two progenitors, and to minor mergers on spheroidal galaxies, i.e. galaxies dominated by their own merger-driven bulge ($M_{\rm{m}} > 0.5 M_{\rm{star}}$).

In the minor merger case, the central galaxy's merger-driven bulge acquires the mass of the satellite, plus the new stars from an eventual burst. 
In the major merger case, the merger remnant is a merger-driven bulge, containing all the stellar mass of the two progenitors, plus the newly formed stars when the cold gas mass available in the collision is bursted (see Lu et al. 2014, Croton et al. 2016). We develop a recipe for the structure of the remnant (or remmant bulge) in analogy with Hatton et al. (2003) and Covington et al. (2011), which we represent in Fig.(\ref{cartoon}) as the \textit{orbital R} radius.

In a merger (both major and minor), the two progenitors spiral in towards the centre of mass, losing orbital energy and momentum to the dark matter halo due to dynamical friction. From the moment the two progenitors reach a distance from each other equal to the sum of their respective radii, energy conservation is assumed and the collision starts.

The total binding energy of the merger remnant is determined by its total mass and radius (Hatton et al. 2003, Covington et al. 2011):
\begin{equation}
E_{\rm{final}} = G \left[ \frac{(M_{\rm{star1}} + M_{\rm{star2}} + M_{\rm{star-new}})^2}{R_{\rm{final}}} \right] ~,
\label{e_final}
\end{equation}
where the mass is the sum of the stellar masses of the two progenitors, and $R_{\rm{final}} = R_{\rm{m}}$ is the stellar half-mass radius of the remnant. $M_{\rm{star-new}}$ is the stellar mass formed during the merger event (from the gas in the progenitors, shocked in the collision):
\begin{equation}
M_{\rm{star-new}} = \epsilon (M_{\rm{gas1}} + M_{\rm{gas2}}) ~,
\label{mstar_new}
\end{equation}
where we assume that there is no gas left in the remnant. The parameter $\epsilon$ is the efficiency of the star formation burst, which depends on the mass fraction of the merger itself and is calibrated on the case of a (1:1) merger (see Croton et al. 2006, Covington et al. 2011):
\begin{equation}
\epsilon = \epsilon_{\rm{1:1}} \left( \frac{M_{\rm{sat}}}{M_{\rm{central}}} \right)^{\rm{\gamma}} ~.
\label{epsilon}
\end{equation}

\noindent Under energy conservation, the remnant energy is equal to the sum of the initial energy of the two progenitors, the total orbital energy at the start of the collision and the energy radiated away from the gas that is forced to condense because of the shocks caused by the collision:
\begin{equation}
E_{\rm{final}} = E_{\rm{initial}} + E_{\rm{orbital}} + E_{\rm{rad}} ~.
\label{e_final_eq}
\end{equation}

\noindent The combined initial gravitational energy of the two progenitors is determined by their total mass in stars and gas and their half-mass radii: 
\begin{equation}
E_{\rm{initial}} = G \left[ \frac{(M_{\rm{star1}} + M_{\rm{gas1}} )^2}{R_{\rm{1}}} +  \frac{(M_{\rm{star2}} + M_{\rm{gas2}} )^2}{R_{\rm{2}}} \right] ~.
\label{e_initial}
\end{equation}

\noindent The collisional energy is given by the orbital energy of the pair of progenitors assuming a circular orbit, calculated at the separation equal to the sum of the two radii $R_{\rm{1}} + R_{\rm{2}}$:
\begin{equation}
E_{\rm{orbital}} = G \left| \frac{(M_{\rm{star1}} + M_{\rm{gas1}} ) (M_{\rm{star2}} + M_{\rm{gas2}} )}{R_{\rm{1}} + R_{\rm{2}}} \right| ~.
\label{e_orbital}
\end{equation}

\noindent Finally, the radiative term accounts for the energy dissipation of the gas component due to shocks, and it depends on the total gas fraction available in the mergers and the total initial energy of the progenitors, with an efficiency parameter $C_{\rm{rad}}$, which Covington et al. (2011) set to $2.75$:
\begin{equation}
E_{\rm{rad}} = C_{\rm{rad}} \cdot E_{\rm{initial}} \cdot \frac{M_{\rm{gas1}} + M_{\rm{gas2}}}{ M_{\rm{star1}} + M_{\rm{gas1}} + M_{\rm{star2}} + M_{\rm{gas2}}} ~.
\label{e_rad}
\end{equation}

\bigskip
\bigskip

We evolve all stellar components (disk, instability-driven bulge and merger-driven bulge) individually, so that thoughout the life of a galaxy, different components come to dominate its evolution at different times. For instance, a spheroidal galaxy can cool gas and form a disk around its merger-driven bulge, and this in turn can develop and instability-driven bulge. This system can later be re-set into a merger-driven bulge by a major merger. Therefore, the relative age of disks and bulges is a direct manifestation of the galaxy assembly history.

\begin{figure}
\includegraphics[scale=0.4]{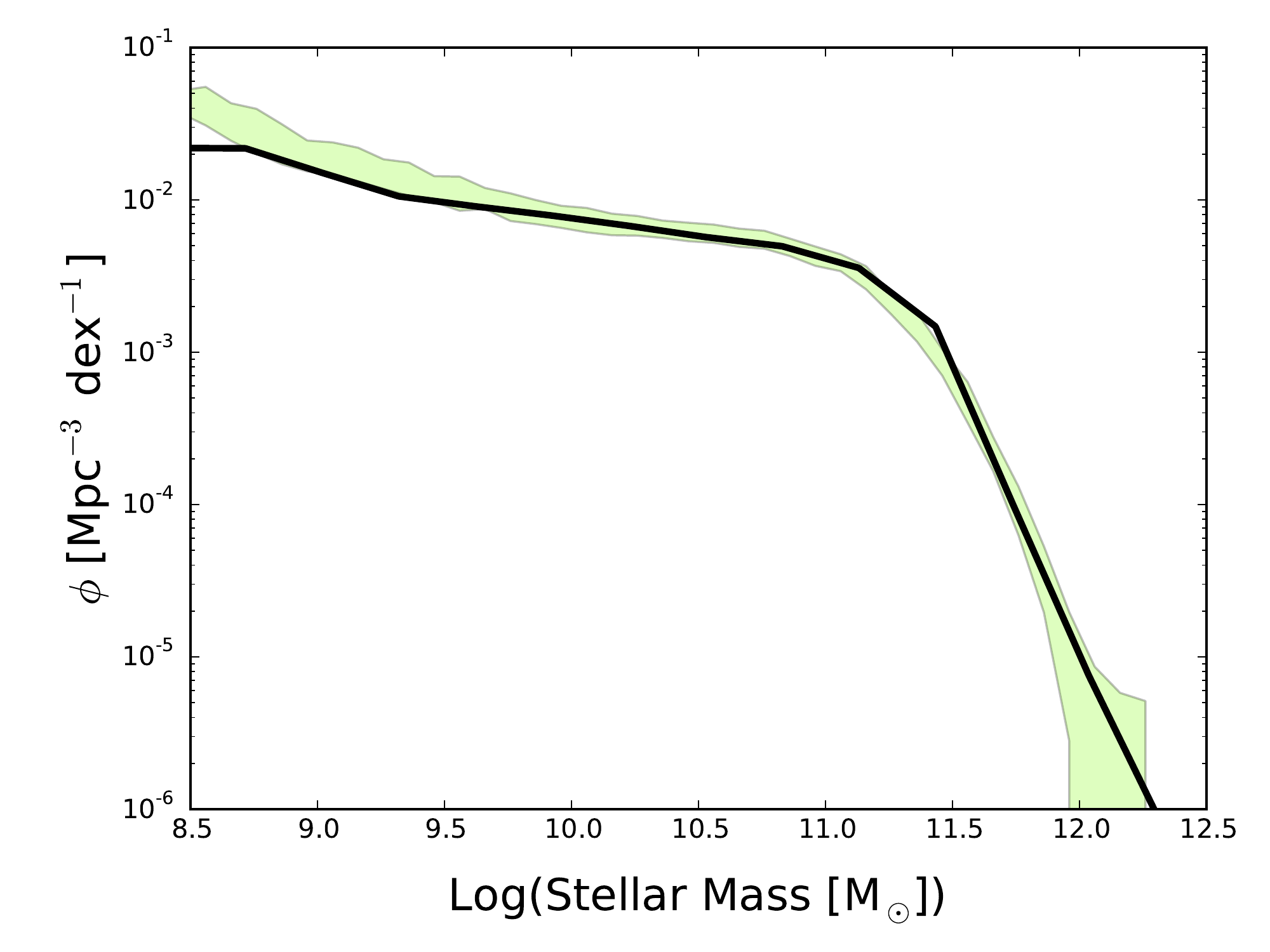}
\caption{The galaxy stellar mass function (\textit{black line}). The shaded area represents data from Drory et al. (2007).}
\label{smf}
\end{figure}

\section{Results}

The model is calibrated using the $z=0$ stellar mass function, represented in Fig.(\ref{smf}) and compared with data from Drory et al. (2007). 
We have introduced a free parameter in the star formation law (Eq.~\ref{SFR}), $\epsilon = 0.25$. Note that this is not directly comparable to the star formation efficiency in SAGE, where a different star formation recipe is used. 
Regardless, the model only needed a minimal recalibration in comparison with SAGE. All the model free parameters have remained the same (see Lu et al. 2014), with the exception of the two parameters connected to supernova feedback, which are affected by the star formation law:
the mass-loading factor due to supernovae, set to $\epsilon_{\rm{disk}} = 1.5$, and the efficiency of supernovae to unbind gas from the hot halo, set to $\epsilon_{\rm{halo}} = 0.15$ (original values: $3.0$ and $0.3$).

Fig.~(\ref{smf}) shows that the model reliably reproduces the stellar mass function down to masses $\rm{Log}(M_{\rm{star}}/M_{\odot}) = 9.0 $. Below this limit, the mass resolution effects of the Millennium simulations cause the stellar mass function to drop away from the data. In the rest of the paper, all results are presented for model galaxies above this mass limit. Unless otherwise indicated, results refer to $z=0$ galaxies.

\subsection{Disk mass-size relations}   

\begin{figure}
\includegraphics[scale=0.4]{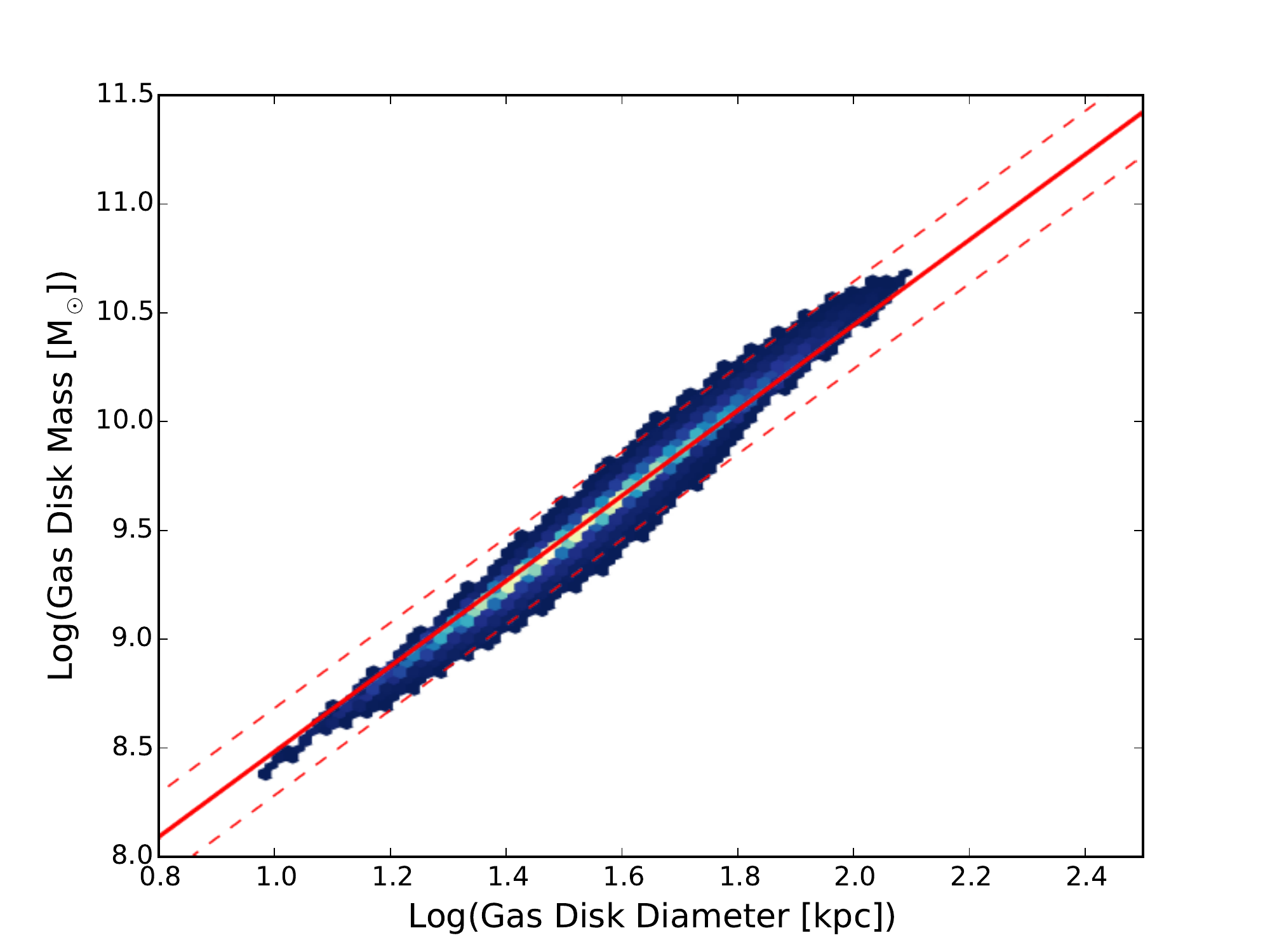}
\caption{The cold gas stellar mass $vs$ diameter for the model galaxies (blue 2D histogram) compared with data from Wang et al. (2014; \textit{red line with scatter}).}
\label{gas}
\end{figure}

To test our angular momentum evolution, Fig.(\ref{gas}) shows the relation between the cold gas mass and diameter for the model galaxies (\textit{blue 2D histogram}). The model predicts a tight relation between gas mass and disk diameter, which spans the entire mass range of the cold gas disks.

Note that the gas exponential density profile of Eq.~(\ref{gas-density}) yields the total gas mass when integrated to infinity, thus we need to truncate the disk when comparing with data. Different definitions of the disk diameter cause the relation to shift along the x-axis, at a fixed slope. We compare the model relation with data of HI disks (\textit{red solid line; dashed lines} represent scatter in the data) from Wang et al. (2014; see also Broeils \& Rhee 1997), where the observed diameter is determined by the HI detection threshold. We find that the slope is very well reproduced, and that the model matches the data when we define the diameter of the gas disk by truncating the density profile at the radius at which the surface density drops to $10 \% $ of its central value.

\begin{figure*}
\includegraphics[scale=0.5]{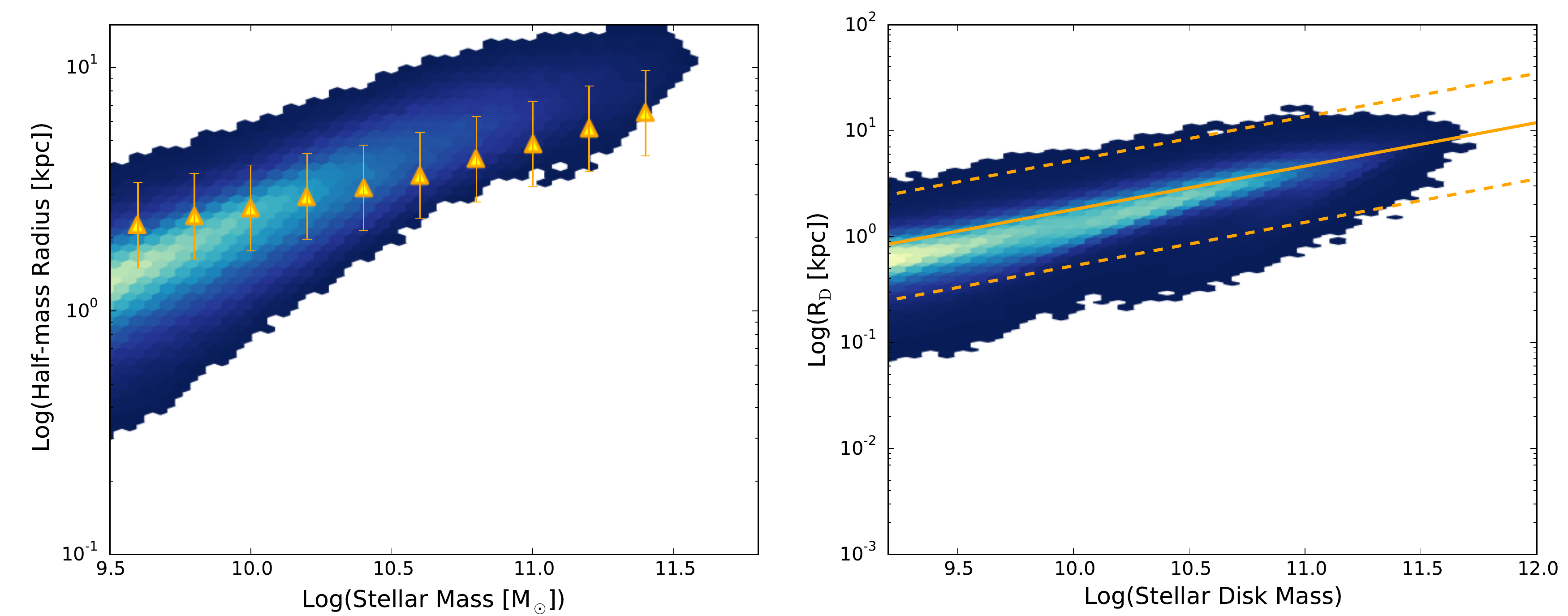}
\caption{\textit{Left panel}: the stellar disk mass $vs$ half-mass radius for the model Sc galaxies, selected as having the stellar disk component accounting for $80 \% $ or more of the total stellar mass (\textit{blue 2D histogram}), compared with data from Shen et al. (2003); see also Guo et al. (2011) for the same comparison. \textit{Right panel}: the stellar disk mass $vs$ scale-length for the model galaxies (\textit{blue 2D histogram}), compared with data of disks from Gadotti et al. 2009 (\textit{orange solid/dashed lines} represent the data and its scatter). }
\label{rstellar}
\end{figure*}

Fig.~(\ref{rstellar}) shows the behaviour of the stellar disk radius as a function of the disk stellar mass. The \textit{left panel} shows the stellar mass $vs$ half-mass radius for the model late-type ($\sim$ Sc) galaxies (\textit{blue 2D histogram}). These have been selected as galaxies for which the stellar disk contains $80 \%$ or more of the total stellar mass (as in Guo et al. 2011). The model is compared with data from Shen et al. (2003), represented by the \textit{orange triangles with errorbars}. 
The \textit{right panel} of Fig.~(\ref{rstellar}) shows instead the stellar mass $vs$ characteristic scale-length for all the stellar disks in the model, regardless of galaxy type (\textit{blue 2D histogram}). The \textit{orange solid/dashed lines} represent the data and scatter of the disk sample of Gadotti et al. (2009).

The model shows good agreement with the Gadotti et al. (2009) data, especially in the slope, thus providing a good sanity check for our angular momentum implementation. 

The agreement is less good with the Shen et al. (2003) data. Although the data sits inside the model scatter, the model slope is steeper than the data one especially at the low mass end, a problem opposite to Guo et al. (2011) when they made the comparison with the same data sample.

Not only is the scatter in the model higher at lower masses, where the stochasticity of the hierarchical assembly has a larger impact on the cooling - star formation - feedback loop, but the overall model performance gets worse at low masses, underestimating the radius. 
This discrepancy can be due to the fact that the model uses analytical recipes for the profiles of the mass components, that are a better representation of massive galaxies, and do not take into account effects more prominent at low masses such as tidal truncations, which would decrease the disk density and increase the radius. In addition, the theoretical recipe does not take into account that at small masses the dark matter to baryons ratio rises. Since the model radius goes as $R_{\rm{D}} \propto 1/V_{\rm{max}}$, the radius drops faster than the stellar mass. Finally, an increased dark matter to baryons ratio would increase the velocity dispersion, with a consequent increase in physical size.

\subsection{Elliptical galaxies scaling relations}

In order to check the reliability of the physical recipes we implement to calculate the merger-driven bulge mass growth and half-mass radius, we test the model galaxies with the Faber-Jackson relation (Faber \& Jackson 1976) and the evolution with redshift of the mass-size relation. 

\begin{figure}
\includegraphics[scale=0.45]{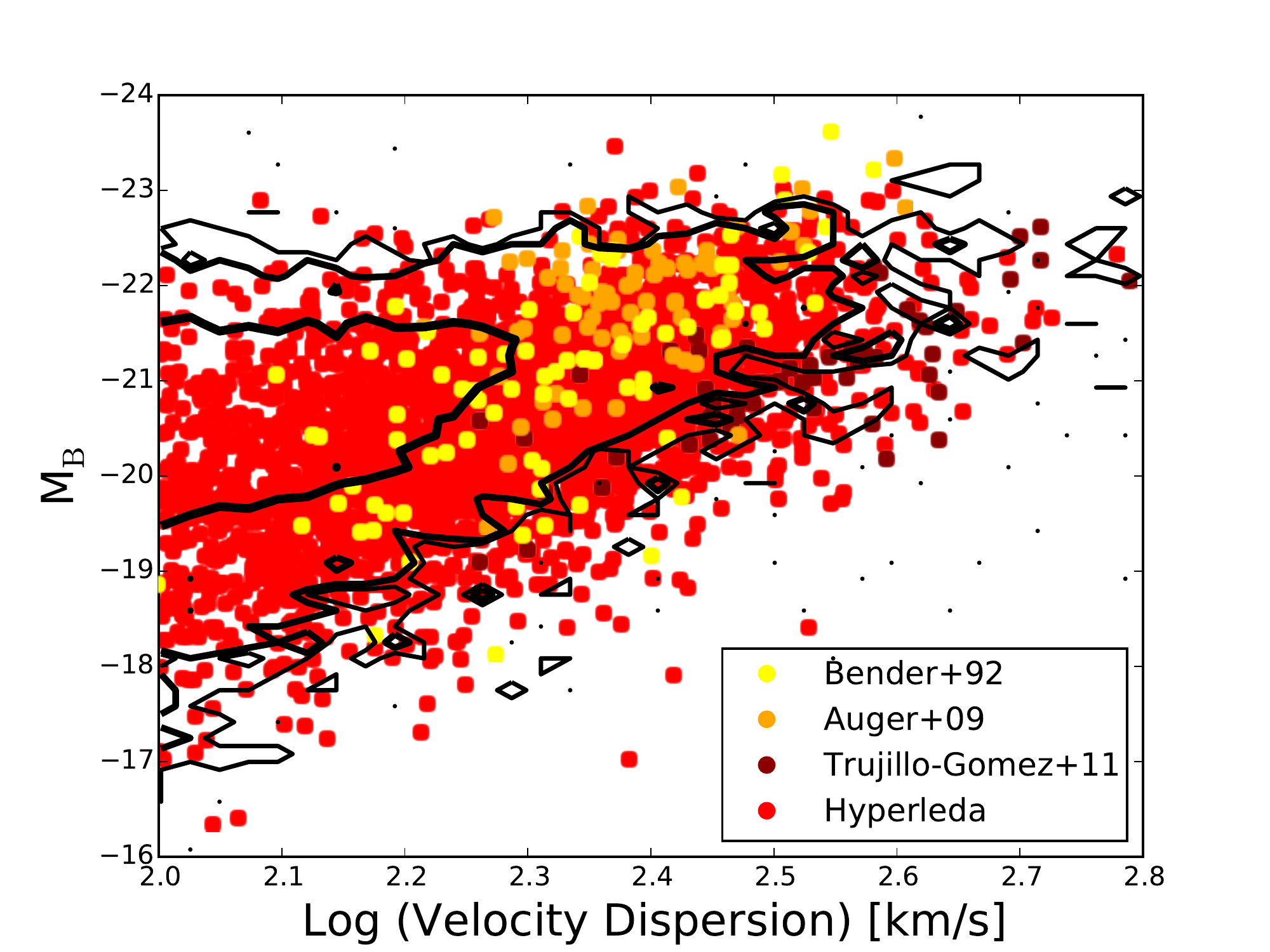}     
\caption{The B-band Faber-Jackson relation for the model elliptical galaxies at $z=0$, represented as the \textit{black contours} for the 68\%, 95\% and 99.7\% confidence levels; coloured circles represent data from the Hyperleda online catalogue, Bender et al. (1992), Auger et al. (2009) and Trujillo-Gomez et al. (2011). }
\label{fj_z0}
\end{figure}

\begin{figure*}
\includegraphics[scale=0.6]{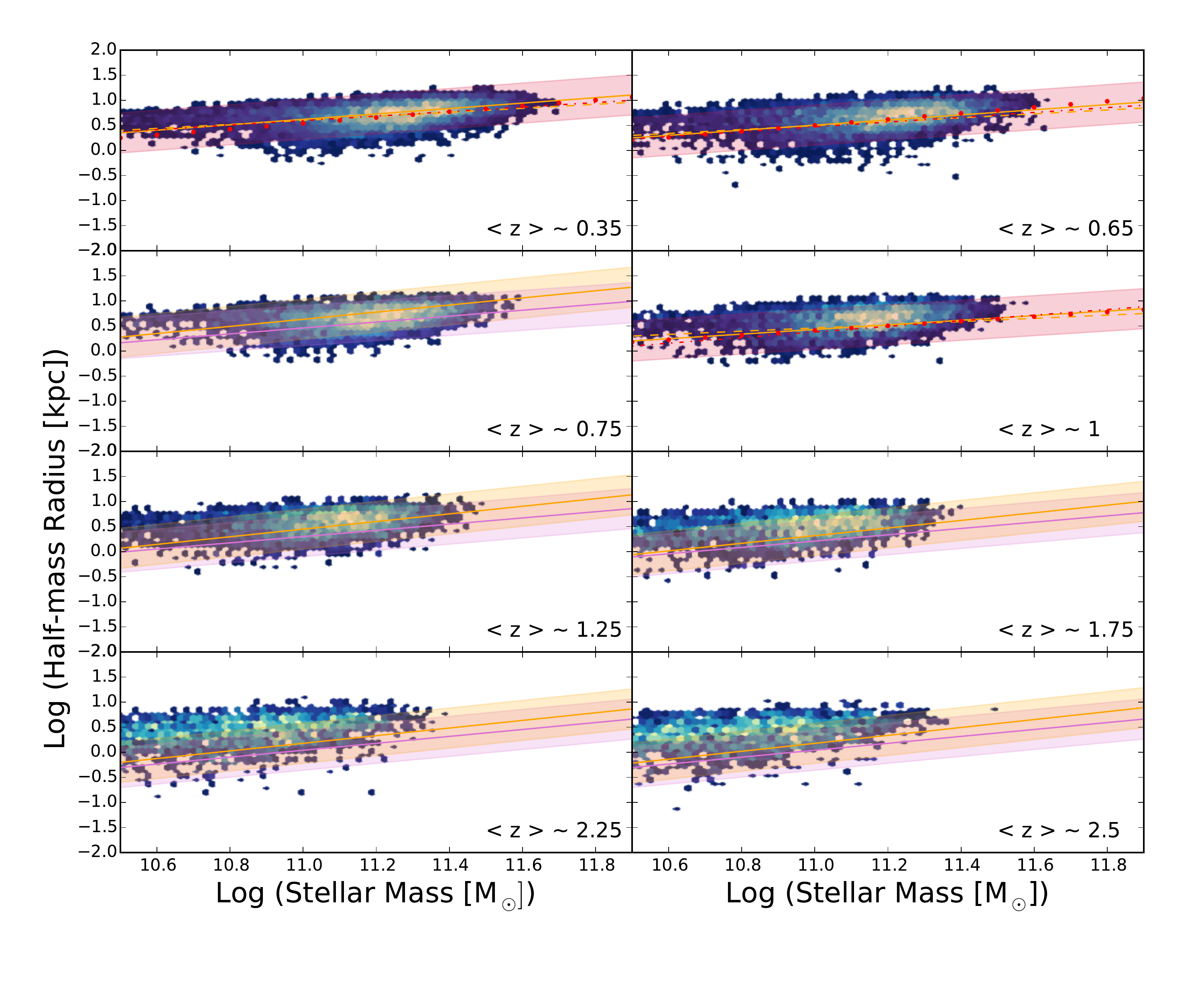}    
\caption{The evolution of the mass-size relation for elliptical galaxies in the model (\textit{blue 2D histogram}), split into 8 redshift bins (as indicated in the panels), compared with datasets from Huertas-Company et al. (2013), Newman et al. (2012), Bernardi et al. (2010 and van der Wel et al. (2014) (\textit{solid lines, with shaded areas representing the $z \sim 0.35$ scatter; see text for description}). }
\label{mass-size1}
\end{figure*}

Figure (\ref{fj_z0}) shows the B-band rest-frame Faber-Jackson relation for the model elliptical galaxies, represented as the \textit{black contours} for the 68\%, 95\% and 99.7\% confidence levels. The model is compared with data from the HyperLeda online catalogue, Bender et al. (1992), Auger et al. (2009) and Trujillo-Gomez et al. (2011). 
The model galaxy spectra and luminosities are computed in post-processing by a spectrophotometric model described in Tonini et al. (2009, 2010, 2012), which uses the galaxy star formation histories from the semi-analytic model and, in this case, the Conroy et al. (2009) synthetic stellar populations models. Luminosities are calculated without dust extinction, as the data have been dust-corrected by the respective authors. The quantity on the x-axis is the line-of-sight velocity dispersion, and we calculate it as follows. 

We assume that merger-driven bulges are akin to elliptical galaxies, which observationally have a de Vaucouleurs surface brightness profile (Kormendy \& Kennicutt 2004). 
Therefore we consider all galaxies in the model with a merger-driven bulge that comprises 90$\%$ or more of the total stellar mass. 
We assume that the stellar mass-to-light ratio is constant with radius (which is expected for merger-driven objects, the structure of which originates from violent relaxation, that erases the previous structure of the progenitors). From the 2D de Vaucouleurs profile, we then obtain a 3D density profile which corresponds to an Einasto profile with index $n=4$ (this is the Sersic index; Kormendy \& Kennicutt 2004, Fisher \& Drory 2010):
\begin{equation}
\rho(r) = \rho_{\rm{0}} \ exp \left[ - \left( \frac{r}{r_{\rm{0}}} \right)^{\rm{1/n}} \right]~.
\label{einasto}
\end{equation}
The characteristic radius $r_{\rm{0}}$ and the central density $\rho_{\rm{0}}$ are determined from the half-mass radius and the total bulge mass. We adopt a NFW (Navarro et al. 1997)
density profile for the dark matter halo. We also assume that both the halo and the bulge are completely pressure supported, so that the gravitational potential as a function of radius is a proxy for the velocity dispersion curve, which takes the form  $\sigma(r) \propto \sqrt{GM(r)/r}$. We calculate the total velocity dispersion curve as $\sigma^2(r) = \sigma_{\rm{halo}}^2(r)+\sigma_{\rm{bulge}}^2(r)$ (see also Tonini et al. 2011, 2014), where $r$ is binned along the density profile up to the half-mass radius $R_{\rm{half}}$. We make this choice to mimic observations, where the velocity dispersion is generally calculated inside one effective radius of the galaxy. 
The observed velocity dispersion is an integrated, luminosity-weighted quantity along the line-of-sight. With the assumption of a constant stellar M/L, we weight the $\sigma$ in every radial bin by the stellar mass contained in that bin, we sum over all bins between $-R_{\rm{half}}$ to $R_{\rm{half}}$, and normalise by the bulge half-mass. This is the final $\sigma$ plotted in Fig.~(\ref{fj_z0}).

The model does a good job in reproducing the Faber-Jackson relation at $z=0$. This implies that both the stellar populations in merger-driven bulges and the dynamical structure of the bulges themselves are well reproduced.

\begin{figure*}
\includegraphics[scale=0.5]{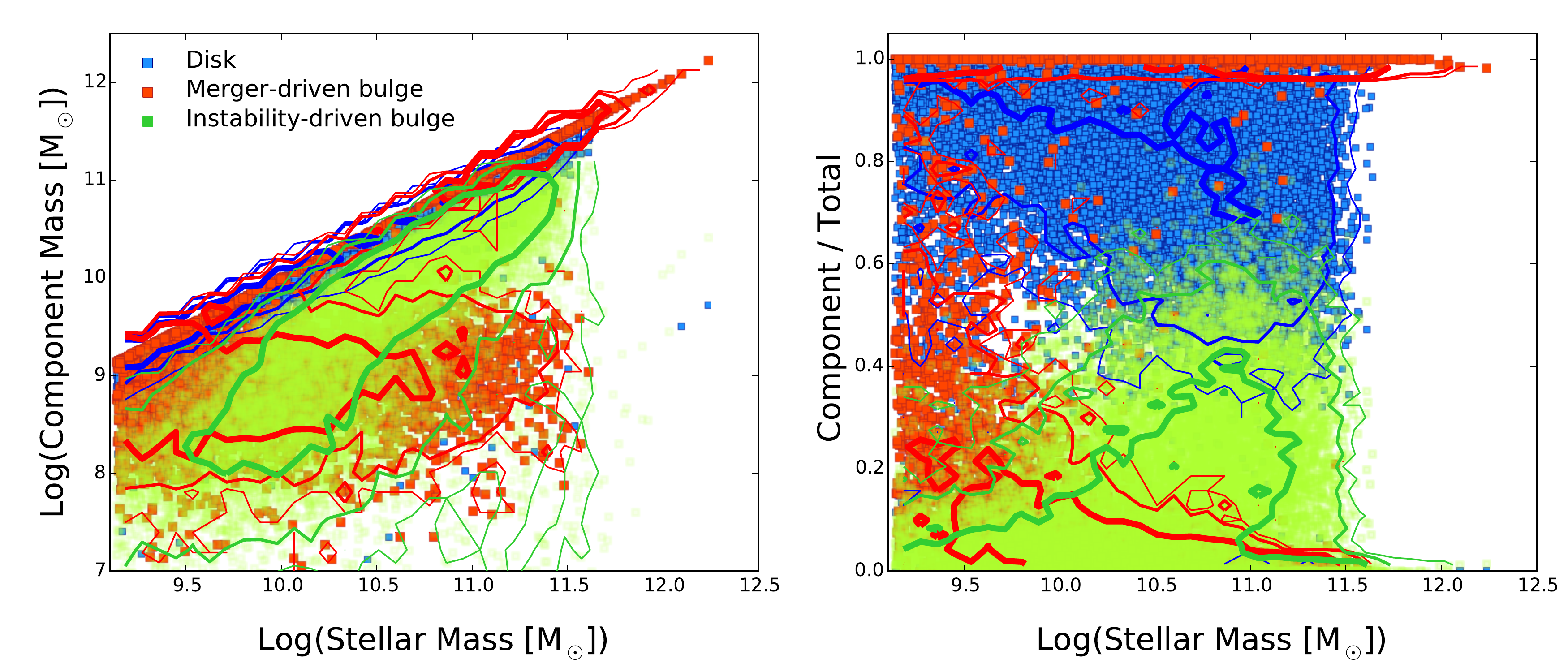}
\caption{\textit{Left panel}: the mass of disks (\textit{blue}), merger-driven bulges (\textit{red}) and instability-driven bulges (\textit{green}) as a function of their host galaxy mass. \textit{Right panel}: the stellar mass ratio of each component (same colour coding) as a function of their host galaxy mass. Contour lines of decreasing thickness represent the 68\%, 95\% and 99.7\% confidence levels.}
\label{bulgemass-stellarmass}
\end{figure*}

The evolution in size of elliptical galaxies has traditionally been difficult to model successfully. 
Figure (\ref{mass-size1}) shows the evolution of the model ellipticals mass-size relation from redshift $z \sim 0.3$ to $z \sim 2.5$ (\textit{blue 2D histogram}), selected as having a merger-driven bulge that account for $60 \%$ at least of the total stellar mass (in analogy with Huertas-Company et al. 2013). 
The model is compared with the following datasets: 1) Bernardi et al. (2010) at redshifts $<z> \sim 0.35, 0.65, 1$ (\textit{red dotted line}); 2) Huertas-Company et al. (2013), where the selection criteria is a bulge-to-total ratio larger than 0.6 and a Sersic index larger than 2.5, at redshifts $<z> \sim 0.35, 0.65, 1$ (\textit{orange solid and dashed lines}); 3) Newman et al. 2012 at redshifts $<z> \sim 0.75, 1.25, 1.75, 2.25, 2.5 $ (\textit{purple lines}); 4) van der Wel et al. 2014 at redshifts $<z> \sim  0.75, 1.25, 1.75, 2.25, 2.5 $ (\textit{orange lines}). In all panels, the shaded areas represent the $z \sim 0.35$ data scatter.  

The mass-size evolution of ellipticals is reproduced well by the model across the redshift range considered. We note that the model has a tendency to produce a somewhat slower size evolution than the data, a discrepancy that starts to become visible at $z > 2$, were it tends to overpredict the radius. 
However it is important to point out that, when making this particular comparison with data, we are forced to compromise on a few aspects. The main caveats are: 

1) the selection of galaxies in the observational samples were made, depending on the data available, based on Sersic index (larger than 2.5), visual early-type morphology, and bulge-to-total ratio following a luminosity profile decomposition.
None of these techniques correspond directly to our selection of the model galaxies, which is based on stellar mass bulge-to-total ratio, and by considering exclusively merger-driven bulges;

2) size measurements and mass measurements based on luminosity and surface brightness profiles are prone to deviate from theoretical estimates due to the inherent uncertainties in the light-to-mass conversion (see for instance Pforr et al. 2012, Tonini et al. 2010, Marchesini et al. 2009), and it is worth noting that any systematic bias in this sense depends on redshift;

3) at high redshift, early-type galaxy stellar mass from observations tends to be overestimated by SED-fitting (see for instance Tonini et al. 2012), due to the intrinsic degeneracies of the technique (in particular between the libraries of star formation histories and stellar population models). 

\subsection{The distribution of bulge types and their properties}

Figure (\ref{bulgemass-stellarmass}) shows the distribution in mass of the three main galaxy components: disks, merger-driven bulges and instability-driven bulges (in \textit{blue, red and green respectively}; the contour lines of decreasing thickness represent the  68\%, 95\% and 99.7\% confidence levels). The \textit{left panel} 
shows the component stellar mass as a function of the total galaxy mass, while the \textit{right panel} shows the stellar mass ratio (mass of component / total stellar mass) as a function of galaxy mass. 

The first thing we notice is that the model produces pure elliptical galaxies through major mergers at all masses from dwarf to giant: these objects form a straight sequence in the mass-mass plot (\textit{left}), and another at mass-ratio equal to 1 in the \textit{right} plot. The vast majority of model galaxies above $\rm{Log}(M_{\rm{star}}/M_{\odot}) \sim 11.5$ are ellipticals.
Merger-driven bulges that do not lie on these sequences form a cloud at lower masses and lower mass ratios; they reach masses of $\rm{Log}(M_{\rm{m}}/M_{\odot}) < 10$ and their mass ratio is mostly below $0.2$, and declining with increasing galaxy mass. These are merger-driven bulges in the centre of disks, they are necessarily dynamically older that the rest of the galaxy and are not evolving. They were formed at the start of the merger tree, and as the hierarchical accretion slowed down, the galaxy had time to form a disk around these objects. 

\begin{figure}
\includegraphics[scale=0.7]{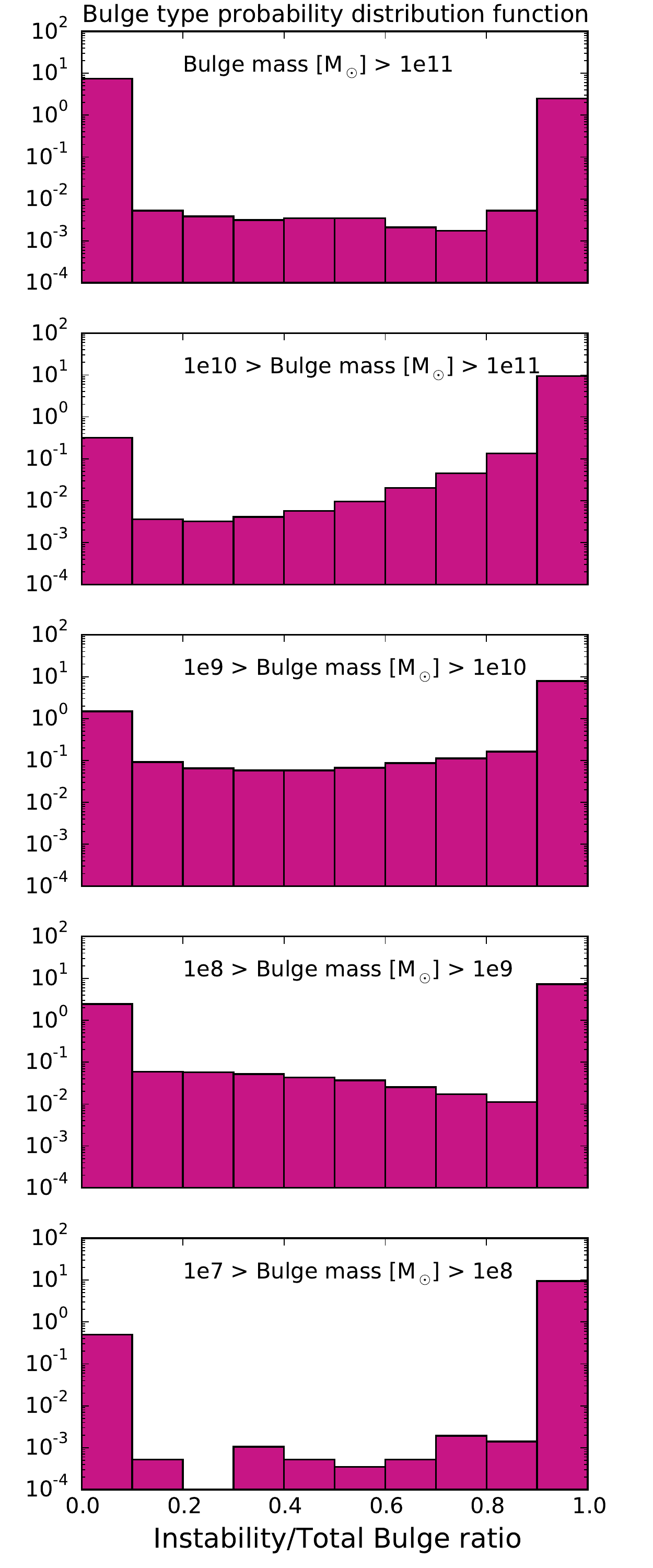}   
\caption{Probability distribution function of the mass ratio between instability-driven component and total bulge mass, for all bulges in the galaxy population: on the \textit{x-axis} a value of 1 indicates a bulge exclusively formed through instability processes, while a value of 0 indicates a bulge exclusively formed through mergers. The \textit{y-axis} represent the number of objects $\rm{Log}(N)$. Each panel represents a bulge mass bin as indicated.}
\label{multiple-ratios}
\end{figure}

The model produces a sequence of pure disks of all masses up to $\rm{Log}(M_{\rm{D}}/M_{\odot}) \sim 12$, although very massive disks are rare. When the disk galaxy has a bulge, the type of bulge depends on galaxy mass. For galaxy masses above $\rm{Log}(M_{\rm{star}}/M_{\odot}) \sim 10$, most bulges are instability-driven, and they tend to grow with increasing galaxy mass, up to $45 \%$ of the total stellar mass, and in some rare cases up to $70\%$. In the same mass regime merger-driven bulges do not live in disk galaxies, but they either comprise all the galaxy mass, or offer a negligible contribution, depending on whether the major merger channel is active or suppressed. In this mass range merger-driven bulges and disks tend to avoid each other in the mass $vs$ mass-ratio space, indicating the alternation of two channels of evolution, violent and quiescent. 
Disk galaxies below $\rm{Log}(M_{\rm{star}}/M_{\odot}) \sim 10$ on the other hand are more likely to host a merger-driven bulge.
This dichotomy echoes the two-stage scenarios for galaxy formation, a fast phase of mass assembly followed by a quiescent one. Evidently for small and isolated overdensities, the envornmnent drives the evolution briefly, and runs out of steam quickly. 

At the low mass end of the galaxy mass distribution, the formation of instability-driven bulges seems to be inefficient, and below $\rm{Log}(M_{\rm{disk}}/M_{\odot}) \sim 10.5$ they amount to less than $20\%$ of the galaxy total mass. 
Above that mass, and in the absence of major mergers, the galaxy population is instead shaped by the competition between disks and instability-driven bulges. These bulges prefer intermediate-to-high masses, because the disk is more active and the galaxy merger tree offers more numerous events that can trigger their growth. In a significant number of galaxies the instability-driven bulge dominates the total mass, turning the galaxy in an object akin to a lenticular (S0).

It is interesting to analyse the relative abundance of merger-driven and instability driven bulges in bins of bulge mass. Fig.(\ref{multiple-ratios}) shows the probability distribution function of bulges in the parameter $M_{\rm{i}}/(M_{\rm{i}} + M_{\rm{m}})$. A value of 1 indicates a galaxy with a bulge grown entirely from instability processes, while a value of 0 indicates a bulge grown entirely from mergers. In the figure, each bulge component has been divided into 5 bins of mass, and we show all bulges, regardless of galaxy type.

This figure shows that instability-driven and merger-driven bulges are two distinct populations, that for the most part do not mix (notice the y-scale is logarithmic). The vast majority of bulges fall in the $0-0.1$ or $0.9-1$ bins, being almost completely instability-driven or merger-driven. At the highest masses, most bulges are merger-driven, and represent the population of giant ellipticals. In the other mass bins, instability-driven bulges are the majority.

Only a small fraction of bulges show a mixed origin, and their numbers become negligible in the highest and lowest mass bins. At the high mass end, a galaxy with a massive instability-driven bulge must have sustained its disk for a long time, with only minor mergers to perturb its growth. This implies that this object lives in a relatively quiet and low-density environment, and it is unlikely that at any time its assembly is dominated by major mergers. On the other hand, major mergers at the high-mass end imply a very dense environment, so even if a massive merger-driven bulge could in principle develop a disk and an instability-driven component, this channel is actually suppressed by environmental conditions. In addition, the time required to reform a disk of comparable mass is very long (as will be discussed later on). At the lowest bulge masses, the assembly history is rather sparse, with a handful of events to shape the galaxy structure. An early fast assembly leads to a merger-driven bulge (with the option to grow a disk in time), while a quiescent assembly leads to a disk, that has time to develop a small instability-driven bulge. However multiple channels, and especially later-time major mergers, are extremely rare, due to the paucity of the merger tree.

Intermediate masses seem to offer a (marginally) more favourable set of circumstances for the formation of mixed bulges, although their probability remains suppressed. At these masses the assembly history is eventful enough to offer a variety of channels of growth, alternating between merger-driven and instability-driven. At the same time, the galaxy mass is not high enough that any configuration is definitive, and there is time for subsequent evolution to develop competing mass components. However, mixed bulges are rare. Environment is the deciding factor, and this plot shows that a significant shift in enviroment is rare (for central galaxies). We expand on this in the Discussion. 
Notice also that at intermediate bulge masses, the probability to have an instability-driven bulge is roughly 10 times higher than a merger-driven one.

\begin{figure*}
\includegraphics[scale=0.6]{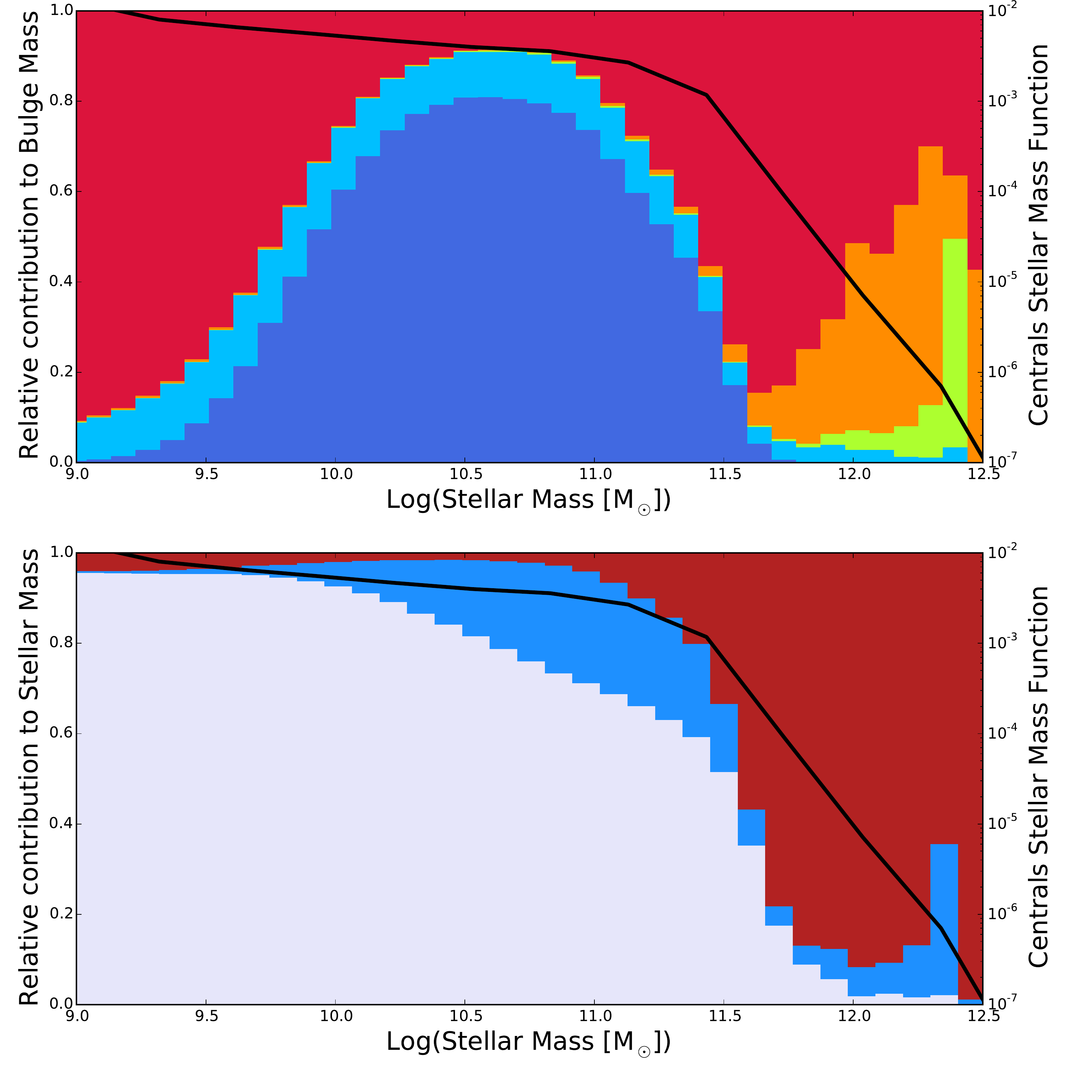}   
\caption{\textit{Upper panel}: relative contribution to the total bulge mass, from the 5 channels of bulge growth depicted in Fig.(\ref{cartoon}), as a function of the total galaxy stellar mass. \textit{Blue}: disk mass that goes into the instability-driven bulge following instabilities of any trigger (including minor mergers); \textit{cyan}: disk mass equal to $\gamma \ M_{\rm{sat, *}}$ accreted by the disk during minor mergers and immediately shedded into the instability-driven bulge; \textit{green} mass accreted during minor mergers that adds to the instability-driven bulge directly, in the case of hybrid morphology (case 3); \textit{orange} mass accreted in minor mergers on central elliptical galaxies, that grows the merger-driven bulge; \textit{red} mass accreted in major mergers, that grows the merger-driven bulge. All masses include the stellar mass produced in starbursts when the bulge growth event includes gas. In this panel disk mass is not depicted. \textit{Lower panel}: relative contribution of disks (\textit{dark blue}), instability-driven bulges (\textit{green}) and merger-driven bulges (\textit{dark red}) to the total stellar mass. \textbf{In both panels the black line} represents the stellar mass function for the central galaxies depicted by the bar plot (right y-axis).}
\label{barplot}
\end{figure*}

The \textit{upper panel} of Fig.(\ref{barplot}) quantifies the relative importance of the different channels of bulge growth depicted in Fig.(\ref{cartoon}), as a function of galaxy mass. For each bin of galaxy total stellar mass, we calculate the total mass in the bulges due to each process, and normalise each bin by the total bulge mass in that bin. The colours represent the following:

\textit{blue}: disk mass that is shed into the instability-driven bulge, following disk instabilities. These include those triggered by minor mergers; in this case the mass that is transferred from the disk into the bulge was already in the galaxy before the merger (Sections 5.1 and 5.2.1);

\textit{cyan}: amount of mass shedded by the disk into the instability-driven bulge after a minor-merger event that brings the disk out of equilibrium, corresponding to $\gamma \ M_{\rm{sat,*}}$ (Section 5.2.1);

\textit{green}: mass directly deposited by minor mergers into instability-driven bulges, in the case the central galaxy is ``bulgy'' but not dominated by a merger-driven bulge (Section 5.2.2);

\textit{orange}: mass deposited into the merger-driven bulge by minor mergers, when the central galaxy is elliptical (i.e. the merger-driven bulge dominates, Section 5.2.3);

\textit{red}: mass deposited in the galaxy by major mergers; in this case the masses of both progenitors grow the merger-driven bulge (Section 5.2.3).

In all cases, we include the stellar mass created during the event by a starburst, if gas in present. We point out that disk mass is not depicted in the upper panel. 

To provide the right perspective on the incidence of the various evolutionary channels on the galaxy population, we also show with the \textit{black line (right y-axis)} the stellar mass function of the galaxies depicted in the bar plot (these are central galaxies only, so the stellar mass function is not the same as that depicted in Fig.~\ref{smf}). 

The \textit{lower panel} of Fig.(\ref{barplot}) shows instead the contribution of disks, instability-driven bulges and merger-driven bulges (\textit{pale grey, blue} and \textit{dark red} respectively) to the total galaxy stellar mass. 

The model predicts that secular processes are responsible for building up the bulge population at intermediate masses, peaking in galaxies with stellar masses between $\rm{Log}(M_{\rm{star}}/M_{\odot}) \sim 10$ and $\rm{Log}(M_{\rm{star}}/M_{\odot}) \sim 11$ (in accord with results from GAMA, Moffett et al. 2016). From the lower panel we see that this range represents the high-mass end of the disk population, whose merger trees have been the most active throughout the galaxy history in producing cooling and star formation, thus creating a conducive environment for instabilities. In the lower panel we notice how the disk distribution is being truncated by the mass seeping into the instability-driven bulge. 

The knee of the galaxy stellar mass function marks the decline of disky galaxies at the high mass end and the dominance of massive ellipticals. At masses around and above $\rm{Log}(M_{\rm{star}}/M_{\odot}) \sim 11.5 $ steady star formation and secular processes cannot compete with the violent environment, and merger-driven evolution dominates. Notice how, for increasing masses in this regime, minor mergers become more and more important in growing elliptical galaxies. 

Below $\rm{Log}(M_{\rm{star}}/M_{\odot}) \sim 10 $ the galaxy population is largely dominated by disks. At such low masses instabilities are sparse. Any activity in the merger tree happens early and is short-lived, and as a consequence bulges in this regime are almost exclusively created in ancient major mergers, and constitute the dwarf ellipticals population or live in the centre of spirals. 

\begin{figure*}
\includegraphics[scale=0.7]{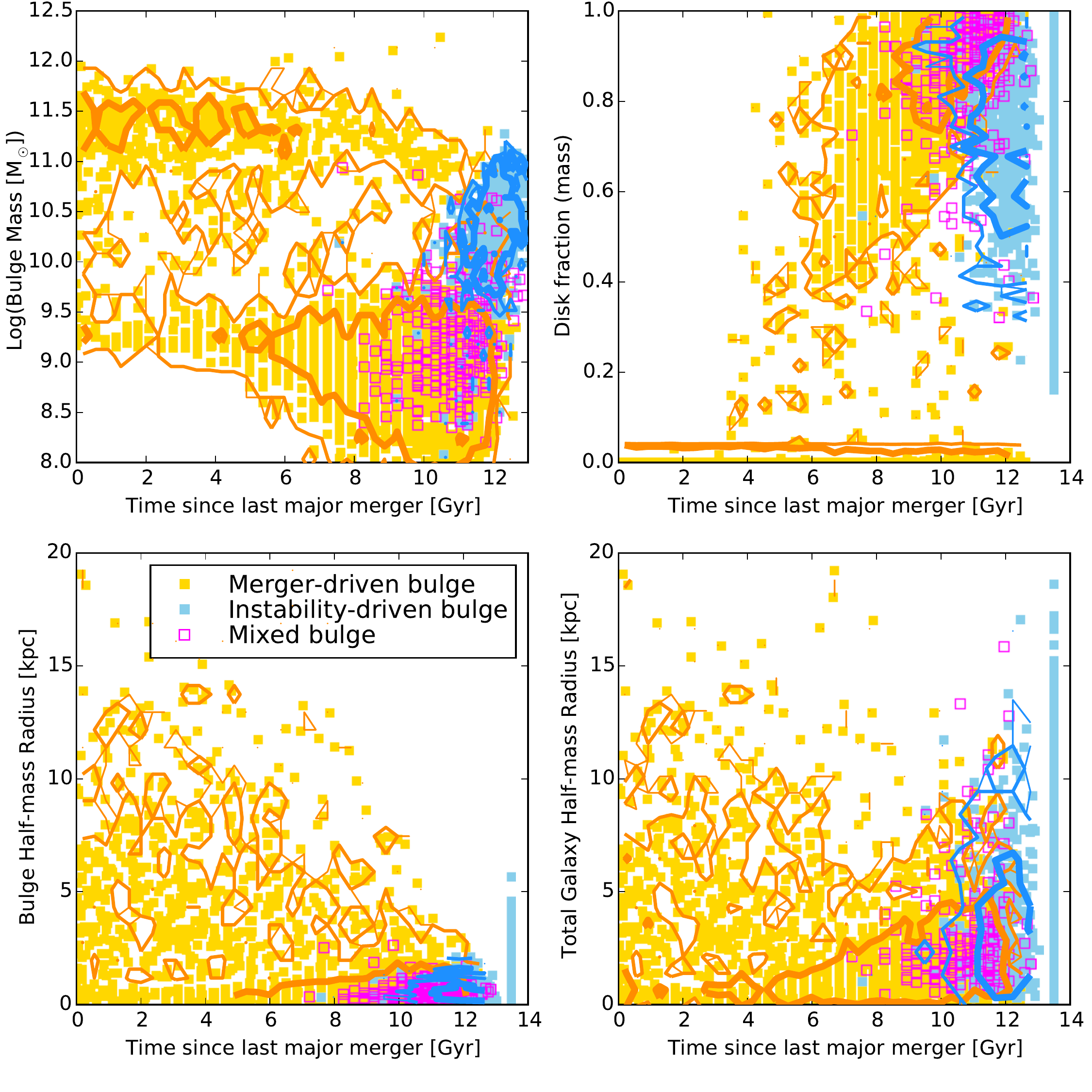}    
\caption{The timescales of bulge formation. \textit{Top left}: the mass of merger-driven bulges and instability-driven bulges as a function of the time since the last major merger occurring in the galaxy. 
\textit{Top right}: disk stellar mass fraction as a function of the total galaxy stellar mass.
\textit{Bottom left}: \textbf{bulge} half-mass radius as a function of the time since the last major merger undergone by the galaxy. \textit{Bottom right}: \textbf{galaxy} half-mass radius as a function of the time since the last major merger. 
Represented are 30,000 central galaxies in a random subsample selected at $z=0$. In all panels the different bulge types are colour-coded as indicated in the legend in the bottom left panel.
The contour lines of decreasing thickness represent the 68\%, 95\% and 99.7\% confidence levels for merger-driven and instability-driven bulges. The narrow vertical strip of blue points at $t \sim 13.5 Gyr$ represents galaxies that have never had a major mergers; for clarity we do not include those in the blue contour lines.}
\label{4panels-mergertimes}
\end{figure*}

Major mergers are ubiquitous as a function of galaxy mass, but the model predicts that secular processes take over in a particular mass regime, dominated by massive disks. This gives us a hint as to the different timescales of formation of the two bulge types, which we quantify in Fig.(\ref{4panels-mergertimes}).
Here we show the behaviour of merger-driven bulges (\textit{orange}) and instability-driven bulges (\textit{light blue}), as well as mixed bulges (\textit{magenta}), in mass and radius as a function of the time passed since the last major merger in the galaxy.
We select the three bulge types as belonging to the ($0-0.1$), ($0.9-1$) and ($0.3-0.7$) bins in Fig.~(\ref{multiple-ratios}) respectively. For plotting clarity, we show a sample of $N=30,000$ model galaxies with a non-zero total bulge mass, selected at random at $z=0$. The contour lines of decreasing thickness represent the 68\%, 95\% and 99.7\% confidence levels for merger-driven and instability-driven bulges.

The \textit{top left panel} of Fig.(\ref{4panels-mergertimes}) shows the bulge mass as a function of the time since the last major merger. 
As major mergers occur at any time, we see a continuous distribution in age of merger-driven bulges. Notice however that they form two distinct sequences in mass. The low-mass objects form for the vast majority at very early times, while the high-mass objects, although more spread out, tend to prefer later times, following the growth of increasingly massive dark matter structures.

Intermediate masses are dominated by the instability-driven bulges, as shown in Fig.(\ref{barplot}). The model predicts that the vast majority of these objects have dynamical ages larger than 8 Gyr, and that there is a tendency for more massive objects to be older, contrary to merger-driven bulges. Mixed bulges are mostly composed of ancient, small merger-driven cores, around which a small instability-driven component is developing. 

Notice that the collisions in major mergers, minor mergers and disk instabilities occur in the model on the same timeframe, corresponding to one simulation timestep. But while
major mergers transform the galaxy structure dramatically in a single episode, minor mergers and instabili- ties produce secular evolution, i.e. composed of many incremental steps. Therefore the build-up of an instability-driven bulge is an effect integrated over time, and it is evident from the \textit{top left panel} of Fig.(\ref{4panels-mergertimes}) that it takes around 8 Gyr (in accord with observational results from Kormendy \& Kennicutt 2004, Bouwens et al. 1999, Fisher et al. 2009). This is a very long time, during which the galaxy assembly history needs to be active in star formation and relatively quiet in merger events. In other words, the perfect conditions for massive disk galaxies.

The \textit{top-right panel} of Fig.(\ref{4panels-mergertimes}) shows the disk fraction, i.e. the stellar disk-to-total mass, as a function of the time since the last major merger. 
The comparison between the contour lines of merger-driven and instability-driven bulges shows 
that most of the instability-driven bulges reach lower disk ratios, meaning that
 secular evolution is more efficient in producing bulges in disk galaxies than fast assembly at early times. 
 The disk fraction for galaxies with instability-driven bulges shows no dependency on the time since the last major merger, indicating that it is primarily the minor merger and instability rate that sets the pace for the bulge growth, rather than the speed of the disk growth\footnote{We point out that the model does not include other internal secular processes in the disk, such as bars for example, the presence of which would accelerate the growth of this type of bulge.}. 

Merger-driven bulges in the \textit{top-right panel} form two distinct sequences. One is for a null disk fraction, across all dynamical ages; these are pure elliptical galaxies, going from dynamically very young to very old (note that ``young'' and ``old'' have no relation to the ages of their stellar populations). A second sequence is formed by merger-driven bulges that live in disk-dominated galaxies, going from dynamically young bulge and low disk fraction to old bulge and high disk fraction. This reflects the growth of disks around merger-driven bulges, and allows us to predict the timescale for disk formation more precisely. For example, if the last major merger happened $\sim$4 Gyrs ago, this plot shows that on average in this time a galaxy can grow a disk that at most amounts to $30\%$ of the stellar mass today. The spread in values of disk fraction for a given lookback time is due to the range in star formation histories (mostly determined by the available cooling of gas).

This panel also shows that merger-driven bulges that are dynamically old are mostly either pure elliptical galaxies or small bulges in disk-dominated galaxies. 
Also, mixed bulges tend to be found in disk-dominated galaxies. 

The \textit{bottom-left panel} of Fig.(\ref{4panels-mergertimes}) shows the half-mass radius of the bulge as a function of the time since the last major merger occurred to the galaxy. When a galaxy has both bulge components, the half-mass radius is calculated as
\begin{equation}
R_{\rm{bulge-half}} = \frac{M_{\rm{m}} R_{\rm{m}} + M_{\rm{i}} R_{\rm{i}}}{M_{\rm{m}} + M_{\rm{i}}}~.
\label{rbulge_half}
\end{equation}

We notice that merger-driven bulges are the largest size-wise, and show a considerable scatter in sizes. Both size and scatter increase for younger dynamical ages, a feature driven by the increase in the maximum mass and size of the merger progenitors with passing time. Instability-driven bulges instead are smaller and more homogeneous in size, with the oldest being the largest. 

There is a narrow strip of objects at a lookback time of 13.5 Gyrs: these are the objects that have never experienced a major merger, and therefore are all instability-driven bulges. Out of the random sample at $z=0$ that we use for this figure, $\sim 90\%$
of galaxies have never experienced a major merger. Out of these $90\%$, less than $15\%$ have an instability-driven bulge. The rest are pure disks, which are not plotted here. This result disagrees with previous claims (see for instance Kormendy \& Fisher 2005 and references therein) 
that pure disks are very hard to produce in the hierarchical galaxy formation scenario.

The \textit{bottom-right panel} of Fig.~(\ref{4panels-mergertimes}) shows the \textit{galaxy} total half-mass radius as a function of the time since the last major merger. Any difference between the two bottom panels of Fig.~(\ref{4panels-mergertimes}) indicates the presence of a disk. 
At lookback times of 6-7 Gyrs or more, disks start to lift the half-mass radius of the galaxy away from the values of the bulge-only radii. This happens for both types of bulges. Since we know that merger-driven bulges are created in major mergers during which disks are destroyed, then disks must form after these types of bulges are already in place. This implies that 6-7 Gyrs is close to the minimum timescale required to form a disk that is dynamically significant compared to the bulge. The higher the lookback time, the more dramatic is the effect of the disk on the galaxy size.

Notice also that, while disks affect the galaxy radius distribution from a lookback time of $\sim$ 6 Gyrs, the instability-driven bulges do not appear for another $2-3$ Gyrs at least. This delay represents the timescale over which disk instabilities and minor mergers manage to grow a significant bulge out of the disk. The speed of instability-driven bulge growth is not limited by the disk growth, but rather by the richness of the merger tree, that determines the number of perturbations and minor mergers and is a proxy for environmental density.

Transformation between bulge types are very localised in the plots of Fig.~(\ref{4panels-mergertimes}). They require an ancient merger-driven bulge, around which a disk is growing to a high disk fraction, and is subject to some degree of instability, at the quiet end of the instability-driven evolution. The behaviour of these bulges in this Figure shows that this sort of merger tree, with early intense activity and a moderate evolution later on, is statistically suppressed by hierarchical clustering.  

In summary, merger-driven bulges form fast and span all dynamical ages, regardless of their masses. On the other hand, the dynamical age of instability-driven bulges is not well defined. On average the more massive the bulge, the longer ago it started its assembly. Environment, which dictates the richness of the galaxy merger tree and the access to gas, plays a crucial part in determining its speed of growth as a function of time.

\subsubsection{Colours}

In this model, hierarchical assembly is the driving force behind bulge growth, whether acting directly with mergers or indirectly by perturbing the disk. In this scenario, the dynamical age of the bulge rarely corresponds to the age of its stellar populations. 

The physical recipes for star formation in the model are based on the instantaneous density of gas and the instantaneous perturbations to the galaxy in the form of mergers and close encounters. The integration of these effects over time constitutes the star formation history of the galaxy, on which its stellar population content depends. It is therefore an important test for the model to check whether the galaxy population is realistic in terms of stellar ages. A fundamental test of this is provided by the colour-magnitude diagram. 

\begin{figure*}
\includegraphics[scale=0.5]{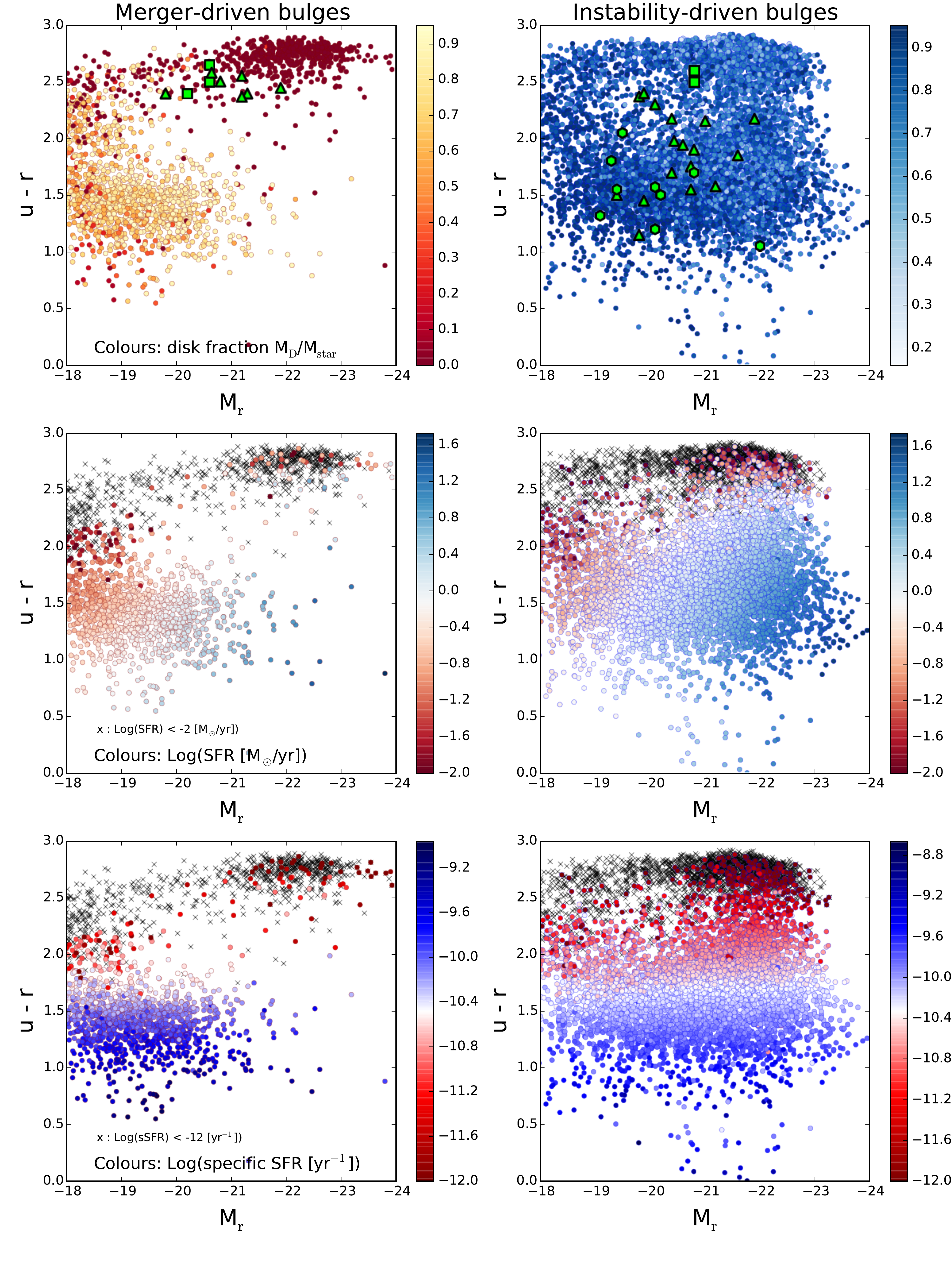}   
\caption{The colour-magnitude diagram $u-r$ $vs$ $M_{\rm{r}}$ for the sample of galaxies depicted in Fig.~(\ref{4panels-mergertimes}), split between galaxies containing merger-driven bulges (\textit{left column}) and galaxies containing instability-driven bulges (\textit{right column}). Model galaxies are colour coded according to 3 different properties as follows; \textit{top row:} disk fraction (stellar mass of the disk $/$ total stellar mass); \textit{middle row:} total star formation rate ($\rm{Log}(SFR [M_{\odot}]/yr$); \textit{bottom row:} specific star formation rate ($\rm{Log} (SFR/M_{\rm{star}} [yr^{-1}])$). Colours in each panel are scaled with the corresponding colourbar; \textit{black crosses} represent galaxies with SFR and specific SFR below the minimum value of the colourbar (as indicated in the relevant panel). 
  The model is compared with $u-r$ $vs$ $M_{\rm{r}}$ data from Drory $\&$ Fisher (2007), depicted by the symbols in \textit{bright green} (portrayed only once in the top panels): they represent galaxies where either a ``classical'' or a ``pseudo'' bulge has been identified; we have associated the former with our merger-driven bulges, and the latter with our instability-driven bulges. Different symbols indicate different galaxy types identified for their hosts: \textit{squares} for early types and S0, \textit{triangles} for spirals from Sa to Sc, and \textit{hexagons} for Sc to Irregulars. }
\label{UR}
\end{figure*}

Figure (\ref{UR}) shows the $u-r \ vs \ M_{\rm{r}}$ (SDSS filters) colour-magnitude diagram for the same sample of galaxies portrayed in Fig.~(\ref{4panels-mergertimes}). We calculate the galaxy photometry with the spectrophotometric model developed in Tonini et al. (2009, 2010), run with the Conroy et al. (2009, 2010) synthetic stellar population spectra . We also made a run with synthetic spectra from Maraston (2005), and found no significant difference in the results for these particular photometric bands. Note that we plot the total galaxy colour and luminosity (as opposed to bulge-only) in order to compare with observational data.
In the \textit{left panels} we show galaxies with merger-driven bulges and elliptical galaxies, while in the \textit{right panels}  we show galaxies with instability-driven bulges (their bulges respectively belong to the ($0-0.1$) and ($0.9-1$) bins in Fig.~\ref{multiple-ratios})). In the 3 rows of panels, we have colour-coded the model galaxies according to 3 different properties, including the stellar disk fraction $M_{\rm{D}}/M_{\rm{star}}$ (\textit{top row}; for plotting clarity we have excluded from this plot galaxies with $M_{\rm{D}}/M_{\rm{star}} > 0.95$ which would saturate the colourbar), the total star formation rate $\rm{Log}(SFR  \ [M_{\odot}/yr])$ (\textit{middle row}), and the specific star formation rate $\rm{Log} (SFR/M_{\rm{star}}  \ [yr^{-1}])$ (\textit{bottom row}). Colours in each panel are scaled with the corresponding colourbar, and \textit{black crosses} represent galaxies with SFR and specific SFR below the minimum value of the colourbar (as indicated in the relevant panel). 

The model is compared with $u-r \ vs \ M_{\rm{r}}$ data from Drory $\&$ Fisher (2007), depicted by the symbols in \textit{bright green} (only once in the top panels). These represent galaxies where either a classical or a pseudo bulge has been identified. We have associated the former with our merger-driven bulges, and the latter with our instability-driven bulges, in accord with the scenario discussed in Drory $\&$ Fisher (2007; see also references therein),
which favours secular processes (instability-driven) for the origin of pseudo-bulges, and mergers for the origin of classical bulges. 
Different symbols indicate different galaxy types identified for their hosts, with \textit{squares} for early-types and S0, \textit{triangles} for spirals from Sa to Sc, and \textit{hexagons} for Sc to Irregulars. The model is able to reproduce the colour and magnitude of the observed galaxies, indicating a good interplay of the dynamical recipes with the star formation histories. 

For galaxies with merger-driven bulges (\textit{left column}), the model predicts a well defined red sequence dominated by elliptical galaxies, spanning the mass range from dwarf to giant and showing a colour-luminosity trend. These galaxies are red and their stellar populations are old, regardless of their large range of dynamical ages. 
These model galaxies also show a very low-to-negligible star formation and specific star formation rates. 
The observed classical bulges from Drory $\&$ Fisher (2007) lie on this sequence of colour and luminosity. 

There are a few bulge-dominated stragglers whose colours have been scrambled bluewards, away from the red sequence. These are likely to be subject to sporadic episodes of star formation, which are rare but can be produced by accretion of gas-rich satellites or gas reincorporation. The specific star formation rate for the majority of these objects is very low, but the low mass-to-light ratio in the $u$ band is such that colours can be significantly affected for a brief time (see Tonini et al. 2012). 

A second population of merger-driven bulges lives in disk-dominated galaxies, which form a small blue cloud at low-to-intermediate masses. These are galaxies that are re-growing a disk around an ancient merger-driven bulge (of age $>$9 Gyrs, from Fig. ~(\ref{4panels-mergertimes})), and their colours are accordingly blue. They have high specific star formation rates, a clear SFR-mass relation and a tendency to be bluer when brighter. These are indications that the mass growth is happening in the disk through in-situ star formation, which is at odds with the conclusions of Fisher $\&$ Drory (2007), who state that if a galaxy contains a classical bulge, then the whole galaxy is on the red sequence, and that the type of bulge (pseudo-bulge $vs$ classical bulge in their case) is as important as the bulge-to-total fraction in determining galaxy colour. However in the luminosity range of the data, most of the model blue cloud contains merger-driven bulges that account for $< 20 \%$ of the total mass, that might just have been missed in the relatively small sample of Fisher $\&$ Drory (2007), which was not designed to be complete. Other larger datasets (for instance Simard et al. 2011) contain galaxies with classical bulges of the same bulge-to-total ratio that agree with the model blue cloud in the upper-left panel.  

Galaxies with merger-driven bulges populate the faint end of the global blue cloud, up to $M_{\rm{R}} \sim -20.5$. Above this magnitude the blue cloud is almost exclusively composed of disk galaxies with a small instability-driven bulge (or no bulge at all) as seen in the \textit{right panels} of Fig.~(\ref{UR}). 

For galaxies with instability-driven bulges, the model predicts a bimodality at all masses and all disk fractions, with well defined blue cloud \textit{and} red sequence. The colours of pseudo-bulges from Drory $\&$ Fisher (2007) also expand up into the red sequence, and as they do so, the galaxy morphology tends to move to earlier types, but only two objects are classified as S0. Model galaxies with instability-driven bulges too tend to go to earlier types when moving from the blue cloud to the red sequence, and in accord with these data, their disk fractions mostly represent galaxies up to Sa types, while S0s are rare.
The spread in behaviour is not as sharp as for merger-driven bulges, however there is a mild tendency that a decreasing disk fraction corresponds to redder colours and decreasing star formation rates.  

The colours show that the model instability-driven bulges live in galaxies with a wide range of stellar ages and star formation rates. 
Both the total and specific star formation rates at fixed luminosity are on average higher for galaxies containing instability-driven bulges than those containing merger-driven bulges. The predicted values of SFR and specific SFR agree with results by Fisher (2006) and Fisher et al. (2009) 
that show how galaxies containing pseudo-bulges, i.e. bulges grown by slow internal processed like disk instabilities, show a wide variety of star formation rates, in most cases indistinguishable from those of the general disk galaxy population.
The blue cloud of galaxies with instability-driven bulges shows a clear mass-SFR relation, and a flat mass-specific SFR relation, with moderate to high levels of star formation. The red sequence is characterised by very low star formation, and the bright end is more active than the faint end. On average a higher disk fraction corresponds to a higher specific star formation rate. 

Notice that the region of intermediate colour between the red sequence and the blue cloud (the so-called ``green valley'') is almost exclusively populated by model galaxies with instability-driven bulges. Elliptical galaxies (\textit{left panels}) might temporarily move into the green valley thanks to bursts of star formation, but there are no disk galaxies with merger-driven bulges that are moving up from the blue cloud into the red sequence. This is because merger-driven bulges are older than the disks, and sit inertly at their centres. 

On the other hand, instability-driven bulges are forming from disk material. 
Disk fraction itself does not seem to be the determining factor, but it is undeniable that something about the growth of the instability-driven bulge can make the galaxy turn red. 
A possible explanation is that the perturbations that cause instability-driven bulges to grow also make the galaxy run out of gas faster. In fact, while the unperturbed galaxy burns through its cold gas reservoire at a steady pace and mantains its blue colours, minor mergers and gravitational instabilities cause bursts of star formation that consume gas with a higher efficiency. The galaxy colours would briefly flash blue, but the subsequent dip in gas density following the burst would interrupt the star formation and cause the colours to turn redder than before the event. The balance between the richness of the (minor) merger history and the rate of steady gas accretion ultimately determines the colours. Since $u-r$ colours are short-lived and the $u$-band mass-to-light ratio is very low, each single episode can alter the galaxy colours considerably, but over time the integrated effect is a turn towards the red. The connection between the growth of the instability-driven bulge and the star formation history will be addressed in future work (Tonini et al. in prep).

The events that trigger the instability-driven bulge growth are more frequent in dense environments and at higher redshifts, with a large variation from galaxy to galaxy. 
Assembly and star formation histories with a lot of activity at very high redshift and low activity at the present time are likely to result in instability-driven bulges living in relatively quiescent disks. Regardless of disk fraction, a mature (i.e. not growing) instability-driven bulge indicates an intert environment and generally low levels of star formation. These objects are akin to the inactive pseudo-bulges discussed in Fisher et al. (2009). Contrary to the morphological quenching scenario, where the growth of the bulge inhibits that of the disk, in the model the disk runs out of steam because the galaxy stops accreting material and a high enough pace to sustain instabilities, and the growth of the bulge is stopped.

\section{Discussion}

Mergers are a characteristic feature of hierarchical clustering. However, major mergers are rare, and
at all redshifts most merger activity is due to small satellites impacting a more massive central galaxy (see for instance Fakhouri et al. 2010). 
However, of the model central galaxies at $z=0$ about $90\%$ never experience a major merger. In their excellent review, Kormendy \& Kennicutt (2004) ask whether secular processes
have had the time to have a significant effect on the galaxy population, given that hierarchical clustering is always active. The answer we can provide is, yes, secular-like processes do not only have the time, but are statistically more favoured than violent processes in shaping the galaxy evolution, and more efficient in building up galaxy bulges. 

Galaxy formation models have traditionally used minor mergers in the same way as major mergers, plunging them to the galaxy centre to grow a one-component bulge, and comparing this to a classical bulge. 
However, the succession of small accretions that comprises most of the merger history of the galaxy is more akin to a semi-steady trickle of material, and the dynamics of each minor encounter is fundamentally different from that of a major merger. 
The physical reasons for this difference are many: 1) the smaller the satellite, the more time it has to lose angular momentum due to dynamical friction, and to align its orbit with the halo spin, thus impacting the disk first; 2) the satellite itself is likely to be disrupted in streams rather than shooting through the centre of the galaxy like a bullet, and therefore is more likely to add its material to the disk; 3) when the disk is the dominant mass component, the incoming satellite is going to feel its gravitational pull much earlier (from larger radii) than that of the bulge (if a bulge is there); 4) the gas in the satellite can be shocked into forming stars before it gets to the actual centre. 

The first operating choice of our model is that minor events of mass accretion do not have the dynamical leverage to alter the existing galaxy structure. In other words, we choose the path of least interference, and let the past galaxy assembly history decide how a minor merger is going to affect the galaxy. This implies that in disky galaxies the disk itself is going to regulate the satellite absorption.
The second operating choice is to create a mass reservoir for secular processes in the disk, distinct from the classical bulge. Observations suggest that mass lost by the disk settles into a flattened bulge-like structure that is kinematically distinct from classical spheroids, it conserves part of its rotation and grows its density over time.

When we apply these two modeling recipes to hierarchical clustering, the consequence is the emergence of a secular channel of evolution alongside the well known merger-dominated one. By following not only the mass but also the angular momentum evolution in the galaxy,  the model we propose is able to produce different speeds of evolution and, crucially, to characterise them with different structural observables.

The speed of mass growth over time and the ratio of smooth accretion vs mergers depend on the richness of the merger tree and are a measure of environment for the central galaxy. We point out that the boundary between smooth accretion and minor mergers is determined by the mass resolution of the simulation. An increase in resolution leads to star formation in smaller halos, thus increasing the minor merger ratio and favouring disk instabilities. A study of mass resolution effects with different N-body simulations is beyond the scope of the present paper, but will be addressed in future work.

The fact that instability-driven and merger-driven bulges very rarely overlap in the same galaxy gives us insight into the statistical behaviour of the merger trees. 
The timescales of secular evolution are very long, and this implies that if two bulges are present, the merger-driven one is ancient. The more recent it is, the shorter the time the galaxy has to grow another component around it (typically these systems can only reach intermediate masses at $z=0$). Hierarchical clustering is fast at early times and slow at late times, and the only way to produce a mixed bulge is to have a fast early tree, which runs out of steam quickly, but lives in a region where gas is still available at later times to grow a disk. Because perturbations grow in time, dense places become denser and empty places become emptier, so that shifts in environment that produce multiple components are suppressed. 
In other words, the conditions for secular evolution vs major mergers are so different, that the fate of a galaxy is decided well beyond each single episode of mass accretion, by environment inside the large scale structure.

The instability-driven bulge is the integrated effect of all the dynamical perturbations and the trickling of small objects into the galaxy. Although the duration of each incremental episode is short (of the order of one dynamical time), 
the overall growth is slow, and dictated by enviromnent. This is a form secular evolution
with external triggers. It is not straightforward to define the dynamical age of an instability-driven bulge, but it is determined by two factors that play against each other, mass and environment. The more massive the bulge is, the earlier its assembly must have started, given the same environmental conditions. On the other hand, for a given mass the bulge in a more active environment is growing at a higher pace, while the one in a poor environment is not growing much and is more mature. 
We note that internal disk processes like bars also contribute to the growth of this type of bulge (see Athanassoula 2005) and complicate this picture, but are not included in the model. 

The balance between fast and slow evolution is imprinted into the galaxy dynamical properties and its stellar populations. 
In accord with Kormendy \& Kennicutt (2004), the model predicts that early-type and late-type galaxies build their dense central components through different evolutionary channels, thus providing a link between morphology and assembly history. 
We can now analyse morphological transformations and photometric properties (a signature of the star formation history) as a function of the speed of mass accretion and environmental effects. For this purpose more data is needed across different photometric bands to refine this analysis. A detailed comparison of the model predictions with observations from IfU surveys like SAMI (Croom et al. 2012, Allen et al. 2015) and MaNGA offers the scope for future work.

An interesting question to ask is whether the instability-driven bulges that have outgrown their disk can be considered early-type galaxies. The model predicts a large range in colour at all masses for these objects, including a significant number of red galaxies. According to photometric criteria, red galaxies dominated by instability-driven bulges are early-types. However these galaxies have disky features too, like a flattened shape and a high rotation velocity. A tantalising possibility is that these objects can be compared to the fast rotators in the ATLAS-3D sample. The relative high number of instability-driven objects compared to merger-driven objects would favour this hypothesis. However, more work is needed to perform this comparison, both theoretically and observationally.

\section{CONCLUSIONS}

We produce a set of physical recipes to follow the mass growth and angular momentum evolution of gaseous and stellar galaxy components, and implement them into a semi-analytic galaxy formation model. We follow gas accretion, star formation, disk instabilities, minor and major mergers, and for each event we evolve the galaxy incrementally, depending on the relative magnitude of the perturbation and the current galaxy structure. We also create a mass reservoir for secular processes, the instability-driven bulge, alongside the merger-driven bulge. The application of this model to hierarchical clustering produces a net divide between the signatures of violent and secular evolution.
Our main results are as follows: 

$\bullet$ The model satisfactorily reproduces the observed mass $vs$ size scaling relations for disks and bulges at $z=0$, the mass-size evolution of elliptical galaxies up to $z \sim 2.5$ and the $z=0$ Faber-Jackson relation. 

$\bullet$ Just a few percent of all galaxies show a mixed bulge, with a merger-driven and instability-driven component. These objects are found at intermediate galaxy masses, and are characterised by an ancient merger-driven bulge, around which a disk has developed an instability-driven component at later times. 

$\bullet$ Secular evolution dominates at intermediate galaxy masses. Most of the instability-driven bulges develop at the high-mass end of the disk population.  Merger-driven bulges dominate at the low-mass end, where they constitute the population of dwarf ellipticals, and bulges in the centre of low-to-intermediate mass disks. They also dominate at the high mass end, constituting the population of giant ellipticals. 

$\bullet$ Minor mergers contribute to the growth of both instability-driven and merger-driven bulges, but they do not represent the dominant channel of mass accretion at any galaxy mass. 

$\bullet$ Disks, merger-driven and instability-driven bulges have vastly different timescales of formation. The timescale of disk formation is of the order of $\sim$6 Gyrs, calculated as the minimum amount of time after the last major merger. Only after the disk is well established can an instability-driven bulge develop, with an average delay of 2-3 Gyrs. Environment plays a major role in regulating its growth. 

$\bullet$ Merger-driven bulges are larger size-wise than instability-driven bulges, and their sizes statistically increase for decreasing redshift of formation.

$\bullet$ We compare the colour-magnitude diagram of galaxies having merger-driven and instability-driven bulges with data, and find good agreement within the observed range. Galaxies dominated by merger-driven bulges lie on the red sequence, where they show negligible star formation. When merger-driven bulges are subdominant in mass, they are dynamically old and only marginally affect the galaxy $u-r$ colour. On the other hand, galaxies with instability-driven bulges span all colour ranges and star formation rates, in agreement with the data. 

$\bullet$ The green valley is almost exclusively populated by galaxies with instability-driven bulges. This suggests that the disk perturbations that lead to the growth of such objects contribute to turn the galaxy colours redwards, possibly by altering the efficiency of the cooling-star formation cycle over time, causing a faster depletion of the gas. 

\bigskip

The mass ratios between disk, merger-driven and instability-driven bulge mirror the galaxy assembly history. In combination with size and angular momentum they provide insight into the timescales of galaxy growth and the transformations of its dynamical structure, and can be used to provide complementary information to that provided by the galaxy stellar populations, to map the galaxy evolution in its entirety.

\section*{Acknowledgments}
We wish to thank the anonymous Referee for their very useful comments and suggestions. CT wishes to thank Michele Cappellari, Julien Devriendt, John Stott, Eric Bell, Laura Wolz and Trevor Mendel for the interesting discussions and helpful suggestions. This work was funded by the Australian Research Council through the ARC DP140103498.

\end{document}